\documentclass[a4paper,dvips,12pt]{article}
\input{axodraw.sty}
\usepackage{amsmath,amssymb,exscale}   
\usepackage{array,multicol}
\usepackage{afterpage,float,flafter}   
\usepackage{epsfig,rotating,pifont,fancybox}   
\makeatletter   
\@addtoreset{equation}{section}   
\makeatother   
   
\def\simlt{\mathrel{\lower2.5pt\vbox{\lineskip=0pt\baselineskip=0pt
           \hbox{$<$}\hbox{$\sim$}}}}
\def\simgt{\mathrel{\lower2.5pt\vbox{\lineskip=0pt\baselineskip=0pt
           \hbox{$>$}\hbox{$\sim$}}}}
\newcommand{\ov}{\overline}   
\newcommand{\Tr}{\mathop{\rm Tr}}

\newcommand{\tr}{\mathop{\rm tr}}

\newcommand{\NPB}[3]{\emph{ Nucl.~Phys.} \textbf{B#1} (#2) #3}   
\newcommand{\PLB}[3]{\emph{ Phys.~Lett.} \textbf{B#1} (#2) #3}   
\newcommand{\PRD}[3]{\emph{ Phys.~Rev.} \textbf{D#1} (#2) #3}

\newcommand{\PTP}[3]{\emph{ Prog.~Theor.~Phys.} \textbf{#1}  (#2) #3}   
\newcommand{\MPL}[3]{\emph{ Mod.~Phys.~Lett.} \textbf{A#1} (#2) #3}

\def\dalemb#1#2{{\vbox{\hrule height .#2pt
        \hbox{\vrule width.#2pt height#1pt \kern#1pt
                \vrule width.#2pt}
        \hrule height.#2pt}}}

\let\a=\alpha    
    
    \let\p=\pi 
     
        \let\Q=\Theta \let\L=\Lambda
 \let\P=\Pi   \let\F=\Phi

 \def\bd{\begin{document}} \def\ed{\end{document}}
\def\ds{\documentstyle} \let\fr=\frac \let\bl=\bigl \let\br=\bigr
\let\Br=\Bigr \let\Bl=\Bigl 
\let\bm=\bibitem
\let\na=\nabla
\let\pa=\partial \let\ov=\overline
\def\ie{{\it i.e.\ }} 
\def\tr{{\mbox{\rm tr}}}
\def\chic#1{{\scriptscriptstyle #1}}
\newcommand{\db}{\rlap{$\partial$}/}
\newcommand{\sla}[1]{\rlap{$#1$}/}
\newcommand{\slap}[1]{\rlap{$#1$}\hspace{0.3ex}/}
\newcommand{\slam}[1]{\rlap{$#1$}\hspace{-0.3ex}/}
\newcommand{\be}{\begin{equation}}
\newcommand{\ee}{\end{equation}}
\newcommand{\beba}{\begin{equation}\begin{array}{lcl}}
\newcommand{\eaee}{\end{array}\end{equation}}
\newcommand{\bea}{\begin{eqnarray}}
\newcommand{\eea}{\end{eqnarray}}
\newcommand{\ba}{\begin{array}}
\newcommand{\ea}{\end{array}}
\newcommand{\td}{\tilde}
\newcommand{\norsl}{\normalsize\sl}
\newcommand{\ns}{\normalsize}
\newcommand{\refs}[1]{(\ref{#1})}
\def\simlt{\mathrel{\lower2.5pt\vbox{\lineskip=0pt\baselineskip=0pt
           \hbox{$<$}\hbox{$\sim$}}}}
\def\simgt{\mathrel{\lower2.5pt\vbox{\lineskip=0pt\baselineskip=0pt
           \hbox{$>$}\hbox{$\sim$}}}}
\def\A{{\cal A}}
\def\a{{\mathcal a}}
\def\V{{\cal V}}
\def\F{{\cal F}}
\def\L{{\cal L}}
\def\p{{\mathcal \phi}}
\def\P{{\cal P}}
\def\Q{{\cal Q}}
\title{   
\vspace*{-0.8cm}   
\begin{flushright}   
 \normalsize{ETH-TH/01-10\\      
IEM-FT-217/01\\
CERN-TH/2001-202\\   
\texttt{hep-th/0108005}}\\ 
 \end{flushright}    
\vspace{1cm}
\Large\textbf{Finite Higgs mass without Supersymmetry}
\author{\large
{\bf I.~Antoniadis~$^1$\footnote{On leave of absence from CPHT, 
Ecole Polytechnique, UMR du CNRS 7644.}, K.~Benakli~$^{1,2}$ and 
M.~Quir{\'o}s}~$^3$\\ \\
\emph{$^1$CERN Theory Division
 CH-1211, Gen{\`e}ve 23, Switzerland }\\
\emph{$^2$Theoretical Physics, ETH Zurich, Switzerland}\\
\emph{$^3$Instituto de Estructura de la Materia (CSIC), Serrano 123,}\\
\emph{E-28006-Madrid, Spain.}}}
\date{}   
\begin{document}
\maketitle
\thispagestyle{empty}
\vspace*{2cm}

\begin{abstract}
We identify a class of chiral models where the one-loop effective potential 
for Higgs scalar fields is finite without any requirement of supersymmetry. 
It corresponds to the case where the Higgs fields are identified with the 
components of a gauge field along  compactified extra dimensions. We present 
a six dimensional model with gauge group $U(3)\times U(3)$
and quarks and leptons accomodated in fundamental and bi-fundamental 
representations. The model can be embedded in a D-brane configuration of 
type I string theory and, upon compactification on a $T^2/{\mathbb Z}_2$
orbifold, it gives rise to the standard model with two Higgs doublets.

\end{abstract}
\vspace{5.cm}   
\date
   
\newpage

\section{Introduction}

In generic non-supersymmetric four-dimensional theories, the mass parameters
of scalar fields receive quadratically divergent one-loop corrections. These
divergences imply that the low-energy parameters are sensitive to 
contributions of heavy states with masses lying at the cut-off scale. 
Such expectations were confirmed by explicit computations in a string 
model in~\cite{ABQ1}.  
In fact, in the case where the theory remains four-dimensional up to
the string scale $M_s \equiv l_s^{-1}$, we found  that the string scale
acts as a natural cut-off: the scalar squared masses are given by a loop
factor times $M_s^2$ and the precise coefficient depends on the details
of the string model.

However, in the case where some compactification radii are larger than the 
string length, which corresponds to the situation where, as 
energy increases, the theory becomes higher dimensional before the string
scale is reached, we found a qualitatively different result.  
There, the one-loop effective potential was found to be finite and
calculable from the only knowledge of the low energy effective field
theory! For instance, in the five-dimensional case with compactification
radius $R > l_s$, we found the scalar squared mass to be given by a loop factor
times $1/R^2$, with exponentially small corrections. The precise factor is now
completely determined by the low energy field theory.

The above behaviour can easily be understood from the
fact that the scalar field considered in~\cite{ABQ1} corresponds to the
component along the fifth dimension of a higher-dimensional gauge field~\cite{Hatanaka}. 
The associated five-dimensional gauge symmetry protecting the scalar field
from getting a five-dimensional mass is spontaneously broken by the 
compactification. As a result a four-dimensional 
mass term of order $1/R$ is allowed and gets naturally generated at one-loop.

In this work we would like to propose a scenario where the Higgs fields  
are identified with the internal components of a gauge field along 
TeV-scale extra-dimensions where the standard model gauge degrees of 
freedom can propagate~\cite{A,AB}. We will not present here a realistic
model for fermion masses; instead, we would like to concentrate on the
main properties of the electroweak symmetry breaking in an example
and postpone a more realistic realization for a future work. 

The adjoint representation of a gauge group containing the standard model 
Higgs, which is an electroweak doublet, should extend the electroweak gauge
symmetry. The minimal extension compatible with the quantum numbers of the 
standard model fermion generations is $SU(3)\times SU(3)\times U(1)$. In
this work, we construct a six-dimensional (6D) model with gauge group
$U(3)\times U(3)$, which can be embedded in a D-brane configuration of
type I string theory. It accomodates all quantum numbers of quarks and
leptons in appropriate fundamental and bi-fundamental representations.
The gauge group is broken to the standard model
upon compactification on a $T^2/\mathbb{Z}_2$ orbifold, leaving as low
energy spectrum the observable world with two Higgs doublets.

We would like to remind that many ingredients were already present in
the literature. For instance, the identification of the Higgs field
with an internal component of a gauge field is not new but 
a common feature of many string models. The use
of this possibility in the case of large extra-dimension scenarios was
already suggested in~\cite{AB}, where two standard model Higgs doublets
were expected to arise from the orbifold action in six dimensions
on $SU(3)$, in a way similar to the model we consider here. 
Moreover, there have been some proposals in various contexts of field theory
where the Higgs field is identified with a gauge field component along 
extra dimensions, leading to finite one-loop mass in the case of smooth 
compactifications~\cite{Hatanaka}. However, a
further essential step was made 
in~\cite{ABQ1} as it was shown that embedding the 
higher dimensional theory in a string framework allows us 
to get a result for one-loop corrections that is calculable in the 
effective field theory. In order to obtain such a result from a field 
theory description it is necessary to assume that the theory
contains an {\it infinite} tower of KK states and not a finite number 
truncated at the cut-off. The absence of ultraviolet divergences in 
the one-loop contribution to the Higgs mass when 
the whole tower of Kaluza-Klein (KK) excitations is taken into account 
has also been discussed by~\cite{savas,Delgado,BHN,gero}. However, in these
cases supersymmetry was necessary in order to cancel the ultraviolet 
divergences in the loop contributions from bosonic (scalar and vector) 
and fermionic fields.

The content of this paper is as follows. In section 2 we derive the
one-loop effective potential for a Higgs scalar identified with a
continuous Wilson line. We show that 
the effective potential is insensitive to the ultraviolet
cut-off in the case of toroidal compactification, and discuss the
requirements in order to remain as such when performing an orbifold
projection. In section 3 we study the minimization of this potential
in the case of two extra dimensions. In section 4 we build a 
model with the representation content of the standard model from a
compactification on a $T^2/\mathbb{Z}_2$ orbifold
of a six-dimensional gauge theory. In section 5 we
compute the one-loop Higgs mass terms for this model reproducing the
results of sections 2 and 3. In section 6 we study the
cancellation of anomalies in our model and obtain the induced
corrections on the effective potential for the Higgs fields. Section 7
summarizes our results and discuss the requirements for more realistic models.
 
\section{The one-loop effective potential}

The  four-dimensional effective potential for a scalar field $\phi$
is given by:
\bea
V_{eff}(\Phi)=\frac{1}{2} \sum_I (-)^{F_I} \int \frac{d^4p}{(2\pi)^4}
\log\left[p^2+M_I^2(\phi)\right] \ .
\label{potential}
\eea
where the sum is over all bosonic ($F_I = 0$) and fermionic ($F_I = 1$) 
degrees of freedom with $\phi$-dependent masses $M_I(\Phi)$. 
In the Schwinger representation, it can be rewritten as:
\bea
V_{eff}(\Phi)&=&-\frac{1}{2} \sum_I (-)^{F_I} \int_0^\infty 
\frac{dt}{t}   \int\frac{d^4p}{(2\pi)^4} \,  e^{-t\left[ p^2+M_I^2(\phi)
\right]}\cr
&=&-\frac{1}{32 \pi^2} \sum_I (-)^{F_I} \int_0^\infty 
\frac{dt}{t^3} \, \, e^{-t M_I^2(\phi)}
\cr
&=&-\frac{1}{32 \pi^2} \sum_I (-)^{F_I} \int_0^\infty 
dl \, \, l   \, \, e^{- { M_I^2(\phi)}/{l}}
\label{potential2}
\eea
where we have made the change of variables $t=1/l$. The integration
regions $t \rightarrow 0$ ($l\rightarrow \infty$) and 
$t \rightarrow \infty$  ($l \rightarrow 0$)  correspond to the 
ultraviolet (UV) and infrared (IR) limits, respectively.

We consider now the presence of  $d$ (large)  extra dimensions compactified on
orthogonal circles with radii $R_i >1$ (in units of $l_s$)
with $i=1, \dots ,d$. 
The states propagating in this space 
appear in the four-dimensional theory as towers of KK
modes of the $(4+d)$-dimensional states labeled by $I$ with masses given by:
\bea
M^2_{\vec{m},I}= M^2_{I}(\phi)+ \sum_{i=1}^{d}
\left[ \frac {m_i+a^{I}_{i}(\phi) } {R_i}\right]^2
\label{mastermass}
\eea
where ${\vec{m}} =\{m_1,\cdots ,m_d \} $ with $m_i$  integers.
In (\ref{mastermass}) the term $M^2_{I}(\phi)$ is
a $(4+d)$-dimensional mass which remains in the limit $R_i \rightarrow
\infty$. The $(4+d)$-dimensional fields 
$\Psi_I$, whose Fourier modes decomposition along the $d$ compact dimensions 
have masses given by
(\ref{mastermass}), satisfy the following periodicity conditions:
\bea 
\Psi_I(x^\mu,  y^i +2 \pi k_i R_i) = e^{i2\pi  \sum_i k_i a^{I}_{i} } 
\Psi_I(x^\mu,   y^i)
\label{periodicity}
\eea
where the $y^i$ coordinates parametrize the $d$-dimensional torus and 
$k_i$ are integer numbers. There are different cases where such a
failure of periodicity appears and generates shifts $a^{I}_{i}$ for
internal momenta. For instance, in the case of a Wilson line, 
$a^{I}_{i} = q^I \oint \frac{dy^i}{2 \pi} g A_i$,
where $ A_i$ is the internal component of a gauge field with gauge coupling 
$g$ and 
$q^I$ is the charge of the $I$ field under the corresponding generator.
Another case is when (\ref{periodicity}) appears as a junction
condition, i.e. as a continuity condition of the wave function, in the
presence of localized potential at $y^i =0$. In this work we will
focus on the first situation.

The cases where $M^2_{I}(\phi)$ are independent 
of $\phi$ are of special interest. Such models, as we shall see
shortly, lead to a finite one-loop effective 
potential for $\phi$. Here, we will consider for simplicity $M^2_{I}=0$, as 
a non-vanishing finite value would otherwise play the role of an 
infrared cut-off but does not introduce new UV divergences.

The effective potential obtained from  (\ref{potential2}) for the spectrum 
in (\ref{mastermass}) with $M^2_{I} =0$ is given by:
\bea
V_{eff}(\phi)|_{\rm torus} =- \sum_I \sum_{\vec{m}} (-)^{F_I} 
\frac{1}{32 \pi^2}  \int_0^\infty 
dl \, \, l \, \, e^{- \sum_i \frac { \left(m_i +a^{I}_{i} 
\right)^2}{ R_i^2 \, l }}
\label{before}
\eea
By commuting the integral with the sum over the KK states, and performing a 
Poisson resummation, the effective potential can be written as:
\bea
V_{eff}(\phi)|_{\rm torus}=-\sum_I (-)^{F_I} \, 
 \frac{\prod_{i=1}^{d} R_i}{32\,  \pi^{\frac{4-d}{2}}}
\sum_{\vec{n}}  e^{ 2 \pi i \sum_i n_i a^{I}_{i}}\, \,   \int_0^\infty 
dl \, \,  l^{\frac{2+d}{2}} 
\,  \,  e^{- \pi^2 l \sum_i n_i^2 R_i^2}
\label{after}
\eea
The term with $\vec{n} = \vec{0}$ gives rise to a (divergent) 
contribution to the cosmological constant that needs to be dealt with in 
the framework of a full fledged string theory. This 
$\phi$-independent part is irrelevant for our discussion and can be forgotten.
For all other (non-vanishing) vectors $\vec{n} \neq \vec{0}$ in (\ref{after}), 
we make the change of variables: $l'= \pi^2\, l\, \sum_i n_i^2 R_i^2$ and 
perform the integration over $l'$ explicitly. This leads to 
a {\it finite} result for the $\phi$-dependent part of the 
effective potential:
\begin{equation}
V_{eff}(\phi)|_{\rm torus} = -\sum_I (-)^{F_I} \, \, 
\frac{\Gamma(\frac{4+d}{2})}{32 \pi^{\frac{12+d}{2}}}\,  \,
\prod_{i=1}^{d} R_i  \,  \,
 \sum_{\vec{n}\neq \vec{0}}  \frac {e^{ 2 \pi i \sum_i n_i a^I_i(\phi)}}  
{\left[ \sum_i n_i^2 R_i^2 \right]^{\frac{4+d}{2}}}
\label{finaltor}
\end{equation}

These results call for a few remarks. 
A generic $(4+d)$-dimensional gauge theory is not expected to be 
consistent and its UV completion (the embedding in a consistent 
higher dimensional theory, as string theory) is needed. However
we found that some one-loop effective potentials
can be finite, computable in the field theory limit and insensitive to most 
of the details of the UV completion under the following conditions:

\begin{itemize} 

\item One of the properties of the UV theory, we made use of, is to
allow to sum over the 
{\it whole infinite} tower of KK modes. This was necessary in order to 
perform the Poisson resummation in (\ref{after}). String theory provides 
an example with such a property. In the string embedding the effective 
potential (\ref{after}) becomes:
\begin{equation}
V_{eff}(\phi)|_{\rm torus}=-\sum_I (-)^{F_I} \, 
 \frac{\prod_{i=1}^{d}\, R_i}{32\,  
\pi^{\frac{4-d}{2}} }\, \, 
\sum_{\vec{n}}  e^{ 2 \pi i \sum_i n_i a^{I}_{i}}\, \,   \int_0^\infty 
dl \, \,  l^{\frac{2+d}{2}} f_s(l)
\,  \,  e^{- \pi^2 l \sum_i n_i^2 R_i^2 }
\label{strings}
\end{equation}
where $f_s(l)$ contains the effects of string oscillators. In
the case of large radii $R_i > 1$, only the
$l\rightarrow 0$ region contributes. This means that the effective
potential receives sizable contributions only from the IR (field theory)
degrees of freedom. In this limit we should have $f_s(l)\rightarrow 1$.  
For example, in the model considered in~\cite{ABQ1}:  
\bea 
f_s(l) = \left[\frac{1}{4 l}  \frac {\theta_2} { \eta^{3}}(il+{\frac{1}{ 2}})
\right]^4 \rightarrow 1 \qquad {\rm for }\qquad l\rightarrow 0, 
\eea
and the field theory result (\ref{finaltor}) is 
recovered~\footnote{Strictly speaking this is true in
consistent, free of tadpoles, models. The known non-supersymmetric
string constructions introduce typically tadpoles that lead to the 
presence of divergences
at some order. However, in the model considered in~\cite{ABQ1} these appear at
higher orders and we were able to extract the finite one-loop
contribution.}.

\item A second, important, ingredient was the absence 
of a $(4+d)$-dimensional mass $M^2_{I}(\phi)$. The effective potential 
contains, for instance, a divergent contribution:
\begin{equation}
\label{potinf}
V^{(\infty)}=\frac{1}{2} \sum_I (-)^{F_I} \, \int\frac{d^{4+d}p}{(2\pi)^{4+d}}
\log\left[p^2+M^2_{I}(\phi)\right]\ .
\end{equation}
While this part identically cancels in the presence of supersymmetry,
in a non-supersymmetric theory it usually gives a contribution
to the $\phi$-dependent part of the
effective potential which is sensitive to the 
UV physics introduced to regularize it. We will consider below
the case where $\phi$ arises as a $(4+d)$-dimensional gauge field. The 
higher dimensional gauge symmetry will then enforce $M^2_{I}(\phi)=0$.

\item  Another issue is related with chirality. Compactification on tori is
known to provide a non-chiral spectrum. Chiral fermions arise in more
generic compactifications as orbifolds. These can be obtained from the
above toroidal compactification by dividing by a discrete symmetry
group. The orbifolding procedure introduces singular points, fixed
under the action of the discrete symmetry, where new localized
(twisted) matter can appear. These new states have no KK excitations
along the directions where they are localized and they generically
introduce, at one-loop, divergences regularized by the UV physics. To
keep the one-loop effective potential finite, we need to impose that
such localized states with couplings to $\phi$ are either absent or
that they appear degenerate between bosons and fermions 
(supersymmetric representations). 

\end{itemize}
The model of \cite{ABQ1}
discusses an explicit string example with the above properties. 

Finally, we would like to comment on higher loop corrections.
UV divergences are expected to appear at two loops, but they must be
absorbed in one-loop sub-diagrams involving wave function
renormalization counterterms. In other words the effect of two-loop
divergences can be encoded in the running of gauge 
couplings~\cite{savas,gero}. Then, requirement of perturbativity imposes
that the string scale should not be hierarchically separated from the
inverse compactification radius (not more than $\sim$ two orders of
magnitude).  An UV sensitive Higgs mass
counterterm is not expected to appear at any order in perturbation theory
because it is protected by the higher dimensional gauge invariance.
On the other hand, in the presence of extra massless localized fields,
there are two-loop diagrams depending logarithmically on the cutoff and 
leading to corrections to the Higgs mass proportional to 
$\log (M_sR)$~\cite{ADD}.

\section{The six-dimensional case}

In this section we would like to study in greater detail the case of
two extra dimensions compactified on a torus. 

\begin{figure}[H]
\centering
\epsfig{file=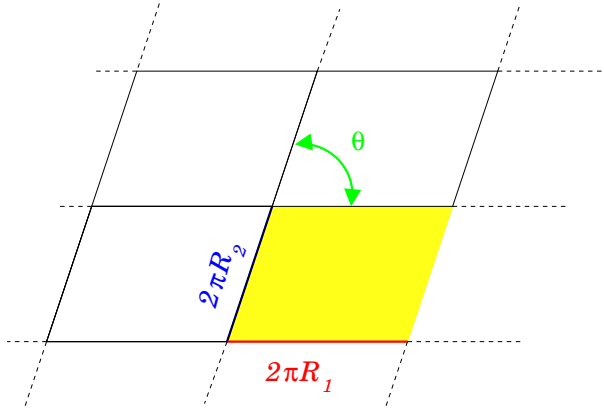,width=.5\linewidth}
\caption{The two-dimensional torus}
\label{torus}
\end{figure}

The torus is parametrized 
by the radii of the two non-contractible cycles $R_1$ and $R_2$ 
and the angle $\theta$ between the directions $x^5$ and $x^6$ 
(see Fig.~\ref{torus}). We will use the notation
$\cos{\theta} = c$, $\sin{\theta} = s >0$. These parameters appear in the
internal metric $G_{MN}$, 
$M,N=5,6$, the torus area $\sqrt{G}$ and the complex structure modulus  $U$
given by:
\bea
G_{MN} = \left(\begin{array}{cc}R_1^2&R_1 R_2 c\\R_1 R_2 c&R_2^2\end{array}
\right); \qquad \sqrt{G}= R_1 R_2 s; \qquad U=\frac {R_2}{R_1} (c+i s)
\label{metric}
\eea
With this notation, the case of orthogonal circles corresponds 
to $\theta=\frac{\pi}{2}$, thus $c=0$. Instead of (\ref{mastermass}), 
the squared mass of the KK excitations now becomes: 
\begin{align}
M^2_{\vec{m},I}&=  \left| \frac {m_2+a_{2}- (m_1+a_{1})U} 
{ \sqrt{Im\,U} G^{1/4}}\right|^2\nonumber\\
&=  
\frac{1}{s^2}\left[\frac {(m_1+a_1)^2}{R_1^2} + \frac {(m_2+a_2)^2}{R_2^2} 
-2  \frac {(m_1+a_1)(m_2+a_2) c}{R_1 R_2}\right]
\label{mass6d}
\end{align}
where we assumed a vanishing six-dimensional mass $M^2_{I}(\phi)=0$.

Plugging the form (\ref{mass6d}) in the effective potential and performing a 
Poisson resummation, one can extract the part of the effective potential 
dependent on $a_1$ and/or on $a_2$ that takes the form:
\begin{equation}
V_{eff}(\phi) = -\sum_I (-)^{F_I} \, \, 
\frac{R_1 R_2 s}{16 \pi^7}\,  \,
 \sum_{\vec{n}\neq \vec{0}}  \frac {\cos{[ 2 \pi (n_1 a_1 +n_2 a_2)}]}  
{\left[  n_1^2 R_1^2 +n_2^2 R_2^2  +2 c n_1 R_1 n_2 R_2\right]^3}
\label{wil6d}
\end{equation}

We consider here  only the case where $a_1$ and $a_2$ are identified with 
Wilson lines:
\bea
a_1=\frac{1}{2\pi}  q \oint g A_5 dx^5 \qquad 
a_2=\frac{1}{2\pi}  q \oint g A_6 dx^6
\eea
where the internal components $A_5$ and $A_6$ of a gauge field have constant
expectations values in commuting directions of the associated gauge groups.
Here $g$ is the gauge coupling and $q$ is the charge of the 
field circulating in the loop.
In such a case the {\it fields} $a_1$ and $a_2$ have no tree-level potential
and the one-loop contribution (\ref{wil6d}) represents the leading order
potential for these fields.

The structure of the minima of the potential (\ref{wil6d}) determines the value
of the compactification radii and torus angle $\cos\theta$ by imposing the
correct EWSB scale at the minimum. 
For instance, in the case of one extra dimension the vacuum expectation value
(VEV) at the minimum uniquely determines the compactification radius. 
It can be easily seen from (\ref{finaltor}) for $d=1$ that the 
minimum of the potential is at $a=1/2$ \cite{ABQ1}. In any
realistic model $a=m_t R$ where $m_t$ is the mass of the fermion which
drives EWSB, i.e. the top in the case of standard model; thus  it follows
that $1/R=2 m_t$ which is the result that was obtained in 
Ref.~\cite{BHN}.

For the case we are considering here, $d=2$, the VEV at the minimum fixes one
of the torus parameters while we have the freedom to fix the other two.
In particular, if we restrict ourselves to the case of equal radii, i.e.
$R_1=R_2\equiv R$, we can still consider the torus angle as a free parameter. 
Using torus periodicity and invariance under the orbifold action we can
restrict the potential to the region $-1/2\leq a_1,\,a_2\leq 1/2$.
In fact the structure of the potential (\ref{wil6d}), 
symmetric with respect to
$|a_2|\leftrightarrow |a_1|$, 
determines that, at the minimum $|a_2|=|a_1|\equiv a$.

\begin{figure}[H]
\centering
\vspace{.6cm}
\epsfig{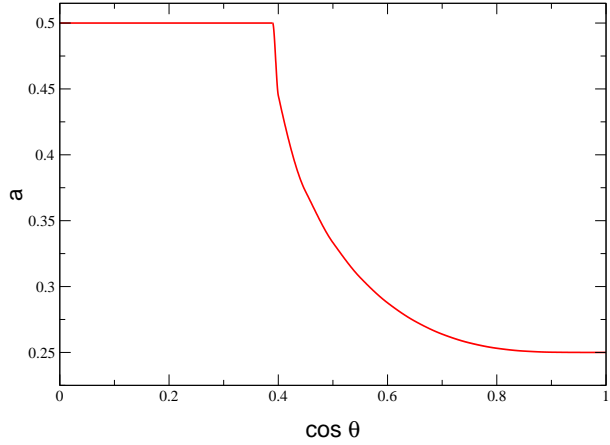}
\caption{The minima of the effective potential (\ref{wil6d}), for $R_1=R_2$,
as a function of $\cos\theta$.}
\label{minima}
\end{figure}

The minimum is plotted in Fig.~\ref{minima} as a function of $\cos\theta$.
We can see that for $\cos\theta<0.4$ the minimum is at $a=1/2$, which 
corresponds to $1/R=2 m_t$. For $\cos\theta>0.4$ the minimum goes from 
$a=1/2$ to $a=1/4$, that would correspond to $1/R=4 m_t\simeq 0.7$ TeV.
Of course in the absence of a tree-level quartic term the corresponding Higgs
mass would be below the experimental bounds and the model becomes 
non-realistic. We will discuss this issue in detail in section 6.

\section{A six-dimensional model}

The $(4+d)$-dimensional Lagrangian for a Yang-Mills gauge field 
$A_{\hat{\mu}}$ coupled to a fermion $\Psi_{(4+d)}$ is given 
by~\footnote{We use the hatted indices  $[\hat{\mu}, \hat{\nu},
\cdots = 0,\cdots, 3, 5, 6, \cdots, 4+d]$  while  $[{\mu}, \nu,
\cdots = 0,\cdots, 3]$ and $[M,N\cdots = 5, 6, \cdots, 4+d]$.}:
\bea
\L = -\frac{1}{2} \Tr F_{\hat{\mu}\hat{\nu}}
F^{\hat{\mu}\hat{\nu}} - i 
\bar{\Psi}_{(4+d)} \Gamma^{\hat{\mu}} D_{\hat{\mu}} \Psi_{(4+d)}
\label{lagran1} 
\eea
where $\Gamma^{\hat{\mu}}$ represent the gamma matrices in
$(4+d)$-dimensions. We use the metric: $\eta_{\hat{\mu}\hat{\nu}}
= {\rm diag}(-1, +1,\cdots, +1) $ and the notation
$F_{\hat{\mu}\hat{\nu}} = \sum_a F^{(a)}_{\hat{\mu}\hat{\nu}} t_a$ and 
$A_{\hat{\mu}} = \sum_a A^{(a)}_{\hat{\mu}} t_a$
where the generators $t_a$ are normalised such that 
$\Tr\left(t_a t_b\right) = \delta_{ab}/2$. With this convention:
\bea
F_{\hat{\mu}\hat{\nu}} &=&
\partial_{\hat{\mu}} A_{\hat{\nu}}  - 
\partial_{\hat{\nu}} A_{\hat{\mu}} -
ig  \left[A_{\hat{\mu}} \, ,\,  A_{\hat{\nu}} \right]\cr
D_{\hat{\mu}} &=& \partial_{\hat{\mu}} + i g A_{\hat{\mu}}
\eea
where $g$ is the tree-level gauge coupling.
Upon toroidal compactification the internal components $A_M$ of the
gauge fields give rise to scalar fields. Some of them will be later  
identified with the  
standard model Higgs field so that the mass structure
given in (\ref{mastermass}) is generated naturally. Furthermore, 
when the scalar fields are identified 
with the internal components $A_M$ of gauge fields, the higher dimensional
gauge symmetry forbids the appearance of a $(4+d)$-dimensional mass term,
i.e. $M^2_{I}(\phi)=0$.

Quartic couplings for the scalar fields are generated from the 
reduction to 4D of the quartic interaction among gauge bosons in 6D
and takes the form:
\bea
V_0 = \frac {g^2}{2}\sum_{M,N=5}^{d+4} Tr \left[ A_M, A_N \right]^2
\label{vzero}
\eea
The tree-level quartic interaction term is absent in the case of 
five-dimensional theory ($d=1$), leading to an unacceptably small Higgs
mass ($\sim 50$ GeV). 
Therefore, a realistic model seems to require $d>1$. We discuss
below the simplest example of $d=2$ extra dimensions.  

We make the following choice of 6D $\Gamma$-matrices~\cite{Fayet}
satisfying the 6D Clifford algebra
$\left\{\Gamma^{\hat\mu},\Gamma^{\hat\nu}\right\}=\, \eta^{\hat\mu\hat\nu}$:
\bea
\Gamma^\mu = \left[\begin{array}{ll}\gamma^{\mu}&0\\0&\gamma^{\mu}\end{array}
\right]; \qquad \Gamma^5 = \left[\begin{array}{ll}0 &-\gamma_{5}\\
\gamma^{5}&0\end{array}
\right]; \qquad \Gamma^6 = \left[\begin{array}{ll}0 &i\gamma_{5}\\
i \gamma_{5}&0\end{array}
\right]
\label{Gammas}
\eea
where $\gamma_{5}$ is the 4D gamma matrix satisfying $(\gamma_{5})^2 =
-1$. We can define the corresponding 6D Weyl projector:
\bea
\P_{\pm} = \frac{1}{2}(1\pm i\Gamma^7)= 
\left[\begin{array}{cc}\frac{1}{2}(1 \mp i\gamma_{5})&0\\
0&\frac{1}{2}(1 \pm i\gamma_{5})\end{array}
\right] 
\eea
so that  $\P_{+}$ and $\P_{-}$ leave invariant the positive and
negative chiralities respectively. The 6D spinor $\Psi_{(6)}$  and the
projectors can be written as: 
\bea
\Psi_{(6)} = \left[\begin{array}{l}\psi_{+}\\\psi_{-}
    \\\Psi_{-}\\\Psi_{+}\end{array}
\right];\qquad 
\P_{+}= 
\left[\begin{array}{llll}1&0&0&0\\0&0&0&0\\ 0&0&0&0 \\ 0&0&0&1\end{array}
\right]; \qquad 
\P_{-}= 
\left[\begin{array}{llll}0&0&0&0\\0&1&0&0\\ 0&0&1&0 \\ 0&0&0&0\end{array}
\right]
\eea
where $\psi_{\pm}$ and their mirrors $\Psi_{\pm}$ are (4D Weyl) two-component
spinors.
The eigenstates of $\P_{+}$ and $\P_{-}$ can be written as:
\bea
\Psi_{(6)+}= \left[\begin{array}{l}\psi_{+}\\ 0
    \\ 0 \\ \Psi_{+}\end{array}
\right] =\left[\begin{array}{l}\psi_L\\ 0
    \\ 0 \\ \Psi_{R}\end{array}
\right] \qquad \Psi_{(6)-}= \left[\begin{array}{l}0\\ \psi_{-}
    \\  \Psi_{-}\\0 \end{array}
\right] =\left[\begin{array}{l}0\\ \psi_R
    \\  \Psi_{L }\\ 0\end{array}
\right] 
\eea
where in the second equality we have dropped the 6D chirality 
indices and used the 4D chirality left (L) and right (R) indices. 

We consider now a six-dimensional theory with gauge group 
$U(3)_3\times U(3)_2$ associated to two different gauge couplings $g_3$
and $g_2\equiv g$ respectively. This model can be embedded in a D-brane
configuration of type I string theory containing two sets of three
coincident D5-branes. The ``color" branes give rise to $U(3)_3 = SU(3)_c
\times U(1)_3$ and contains the $SU(3)_c$ of strong interactions.
Similarly, the ``weak" branes give rise to $U(3)_2 = SU(3)_w \times
U(1)_2$ where $SU(3)_w$ contains the weak interactions. This is the
smallest gauge group that allows to identify the Higgs doublet as
component of the gauge field. Indeed, the adjoint representation of
$SU(3)_w$ can be decomposed under $SU(2)_w\times U(1)_1$ as
\bea
{\bf 8}={\bf 3}_0+{\bf 1}_0+{\bf 2}_3+{\bf {\bar 2}}_{-3} \, ,
\eea
where the subscripts are the charges under the $U(1)_1$ generator
$\Q_1=\sqrt{3}\lambda_8$ with gauge coupling $g/\sqrt{12}$. 
We chose the normalization  of the generators $\Q_2$ and $\Q_3$ of 
$U(1)_2$ and $U(1)_3$ such that the
fundamental representation of $SU(3)_i$ has $U(1)_i$ charge 
unity~\cite{AKT}. The corresponding gauge couplings are then given by
$g/\sqrt{6}$ and $g_3/\sqrt{6}$, respectively.

In addition to the gauge fields, the model contains three families
of matter fermions in the representations 
\bea
L_{1,2,3} =({\bf 1},{\bf 3})^+_{(0, 1)}, \qquad 
D^{\, c}_{1,2,3} =({\bf {\bar 3 }}, {\bf 1})^+_{(-1,0)}, 
\\
Q_{1} =({\bf 3},{\bf {\bar 3 }})^+_{(1,-1)} \qquad 
Q_{2} =({\bf 3},{\bf {\bar 3 }})^-_{(1,-1)} \qquad 
Q_3 =({\bf {\bar 3 }}, {\bf 3})^-_{(-1,1)} 
\label{repr}
\eea
where the notation $({\bf r_3}, {\bf r_2})^{\epsilon}_{(q_3,q_2)}$ 
represents a
six-dimensional Weyl fermion with chirality $\epsilon = \pm$ in the
representations ${\bf r_3}$ and ${\bf r_2}$ of $SU(3)_c$ and $SU(3)_w$, 
respectively,
and $U(1)$ charges $q_3$ and $q_2$  under the generators $\Q_3$ and
$\Q_2$. The choice of the quantum numbers ensures the absence of all
irreducible anomalies in six dimensions (see section 6). 

In a D-brane configuration, the states $Q_{i}$ arise as fluctuations of 
open strings stretched between the color and weak branes.
In contrast, the open strings giving rise to  $L$ and  $d^{\,c}$ 
need to have one end elsewhere as $L$ and $d^{\,c}$ carry charges
only under one of the $U(3)$ factors. This requires 
the presence of another brane in the bulk, where we
assume that the associated gauge group  
is broken at the string scale and is not relevant for our discussion.
The details of the derivation of this model are presented in appendix A,
along with two alternative possibilities of quantum number
assignments that we do not use in this work.

As the six-dimensional chiral spinors contain pairs of left and right 4D
Weyl fermions, the 6D model contains, besides the 
standard model states, their mirrors. Thus, the leptons appear as:
\bea
L_{L} = \left(\begin{array}{l} l  \\ {\tilde e}\end{array} \right)_L 
\qquad \qquad {\rm and \, }\qquad \qquad 
L_{R} = \left(\begin{array}{l} {\tilde l}  \\ e\end{array} \right)_R 
\eea
while the quark representations are: 
\bea 
Q_{1,2L} = \left(\begin{array}{l} q  \\ {\tilde u}\end{array} \right)_L 
\quad 
Q_{1,2R} = \left(\begin{array}{l} {\tilde q} \\ u\end{array} \right)_R
\quad 
Q_{3L} = \left(\begin{array}{l} {\tilde q}^c  \\ u^c\end{array}
\right)_L, 
\quad 
Q_{3R} = \left(\begin{array}{l} q^c  \\ {\tilde u}^c\end{array} 
\right)_R
\eea
where $q$, $l$ are the quark and lepton doublets, and $u^c_L$,  $d^c_L$, 
$e_R$ their weak singlet counterparts,
while ${\tilde q}$, ${\tilde l}$, ${\tilde u}$ and  ${\tilde e}$ are
their mirror fermions.

To obtain a chiral 4D theory from the six dimensional model, we perform
a $\mathbb{Z}_2$ orbifold:  
\bea 
x^5 \rightarrow -x^5  \qquad x^6 \rightarrow -x^6 
\eea 
Each state can be represented as $|gauge> \otimes
|spacetime>$ where $|gauge>$ represents the gauge quantum numbers
(singlet, fundamental or adjoint representation of $U(3)$'s), while
$|spacetime>$ represent the spacetime ones  (scalar, vector or
fermion). The orbifold acts on both of these quantum numbers.

The orbifold action on the spacetime quantum numbers is chosen to be:
\bea
\begin{array}{llll}{\rm even \, :}&A^\mu \rightarrow A^\mu 
  &\qquad \qquad {\rm odd \, :}&  A^{M} \rightarrow -A^{M}
\end{array}\qquad M=5,\, 6
\eea
The adjoint of $SU(3)$ is represented by $3\times 3$
matrices $t_a = \frac{\lambda_a}{2}$, where $\lambda_a$ are the
well-known Gell-Mann matrices. The orbifold action on the adjoint
representation of $U(3)_2$ is defined by:
\bea
t_a \rightarrow
\Theta^{-1} t_a \Theta \, \qquad \qquad {\rm with \, } \qquad
\Theta= \left(\begin{array}{lll}\, 1&\, 0&\, 0\\
\, 0&\, 1&\, 0\\\, 0&\, 0&-1 
\end{array}
\right)
\eea

As a result of combining the two actions, the invariant states from
the adjoint representation of $U(3)_3$ are the 4D
gauge bosons, while from the adjoint representation of $U(3)_2$ we
obtain the $U(2)\times U(1)$ gauge bosons $A_{\mu} = \sum_{a=1,2,3,8}
A^{(a)}_{\mu}  \frac{\lambda_a}{2} + A^{(0)}_{\mu} \cdot 
\frac{\bf 1}{\sqrt{6}}$:
\bea
 A_{\mu} =  \frac{1}{2} \left(\begin{array}{ccc}\, W_3 +
\frac {1}{\sqrt {3}}\, A^{(8)}+  {\sqrt {\frac{2}{3}}}
A^{(0)}&\, 
{\sqrt {2}}{W^+}&\, 0\\\,
{\sqrt {2}}{W^-}&\, - W_3+\frac {1}{\sqrt {3}}\, A^{(8)}
+{\sqrt {\frac{2}{3}}} A^{(0)}&\, 
0\\ \,
0&\, 0& \frac {-2}{\sqrt {3}}\, A^{(8)} + {\sqrt {\frac{2}{3}}}A^{(0)}
\end{array}
\right)_{\mu}
\nonumber \\
\label{gauge1}
\eea
as well as the scalar fields  $H_{M} = \sum_{a=4,5,6,7}
A^{(a)}_{M}  \frac{\lambda_a}{2}$ where $M=5,6$. This takes the form:
\bea
 H_{M} = \frac{1}{2}
\left(\begin{array}{ccc}\, 0&\,0&\, A^{(4)}_{M}+ i A^{(5)}_{M} \\
\, 0&\,0&\, A^{(6)}_{M}+ i A^{(7)}_{M} \\
\, A^{(4)}_{M}-i A^{(5)}_{M}&\, A^{(6)}_{M}- i A^{(7)}_{M}&\,0 \end{array}
\right)  = \frac{1}{2}
\left(\begin{array}{lll}\, 0&\,0&\, H_M^+ \\
\, 0&\,0&\,  H_M^0\\
\, H_M^- &\,  H_M^{0*}&\,0 \end{array}
\right) 
\nonumber
\eea
It is useful to define $H=H_{5}-\gamma_5 H_{6}$ :
\bea
H=  \frac{1}{2}
\left(\begin{array}{ccc}\, 0&\,0&\, H_5^+ -\gamma_5 H_{6}^+ \\
\, 0&\,0&\,  H_5^0-\gamma_5 H_{6}^0 \\
\, H_5^- -\gamma_5 H_{6}^- &\,  H_5^{0*}-\gamma_5 H_{6}^{0*} &\,0 \end{array}
\right)=  
\left(\begin{array}{ccc}\, 0&\,0&\, H_2^+  \\
\, 0&\,0&\,  H_2^0 \\
\, H_1^-  &\,  H_1^{0} &\,0 \end{array}
\right)
\label{Hdef} 
\eea
In addition, the orbifold projection acts on the
fermions in the representation $r_f$ of $U(3) \times U(3)$ as:
\bea
\begin{array}{lllll} r_f \rightarrow \Theta  r_f:&  
 (1,3)_L & (3,{\bar 3})_L&({\bar 3},3)_R \nonumber \\
r_f \rightarrow -\Theta r_f: &      (1,3)_R & 
(3,{\bar 3})_R&({\bar 3},3)_L \nonumber \\
r_f \rightarrow  r_f:
&({\bar 3},1)_L &&& \nonumber \\
r_f \rightarrow - r_f:
 & ({\bar 3},1)_R  & &&
\end{array}
\eea
leaving invariant, in the massless spectrum,
just the standard model fields and projecting the mirrors away. 

The model contains three $U(1)$ factors corresponding to the generators
$\Q_i$ with $ i= 1,2,3$. As we will discuss in more details 
in section 6, there is only one anomaly free linear combination
\bea
\Q_Y &=&\frac{\Q_1}{{6}} -  \frac{2 \Q_2}{3} -\frac{\Q_3}{3}  
\label{hypercharge}
\eea
identified with the standard model hypercharge. The corresponding gauge
coupling is given by:
\bea
\frac{1}{g_Y^2}= \frac{3}{g^2}+ \frac{2}{3}\frac{1}{g_3^2}
\eea
which corresponds to a weak mixing angle $\theta_w$ (at the string scale)
given by:
\bea
\sin^2{\theta_w}=\frac{1}{4 + \frac{2}{3}\frac{g^2}{g_3^2}}
\eea
This relation coincides with one of the two cases considered in
Ref.~\cite{AKT}, which are compatible with a low string scale. Of course,
a detailed analysis would need to be repeated in our model to take into
account the change of the spectrum above the compactification scale.

The other two $U(1)$'s are anomalous. In a consistent string theory 
these anomalies are canceled by appropriate shifting of two axions. 
As a result the two gauge bosons become massive, giving rise to two 
global symmetries. One of them corresponding to $\Q_3$ is the ordinary 
baryon number which guarantees proton stability.

The projection on the fermions as chosen above leaves invariant 
only the standard
model representations and projects away the mirror fermions from the
massless modes. The low energy spectrum is then the standard model one 
with two Higgs doublets $H_1$ and $H_2$ as defined in (\ref{Hdef}).

The Higgs scalars have a quartic potential at tree level given in
(\ref{vzero}). As a function of the neutral components of the 
fields $H_1$ and $H_2$ the potential is given 
by~\footnote{We will see in section 6 that this potential gets corrected 
due to the presence of $U(1)$ anomalies.}:
\bea
V_0(H_1^0,H_2^0)= \frac{g^2}{2} \left( |H_1^0|^2 - |H_2^0|^2 \right)^2
\label{vtree}
\eea
which corresponds to the one of the minimal supersymmetric standard model 
with ${g'}^2 = 3 g^2$ due to the embedding of the hypercharge generator 
inside $SU(3)_w$ as given by (\ref{hypercharge}).

The Higgs field coupling to fermions is given by:
\bea
- i \bar{\Psi}_{(4+d)} \Gamma^{M} D_{M} \Psi_{(4+d)}\rightarrow 
\bar{\Psi}_{(4+d)} \Gamma^{M} \left[ -i \partial_{M} + g \sum_{a=4,5,6,7}
A^{(a)}_{M}  \frac{\lambda_a}{2}
\right] \Psi_{(4+d)} 
\label{fer1}
\eea
and leads to generation of fermion masses when the Higgs fields
acquire VEVs.

\section{One-loop Higgs mass}

The Higgs scalars $H_5$ and $H_6$, or equivalently $H_1$ and $H_2$, 
arise as zero modes in the dimensional reduction of the six-dimensional
gauge field on the torus. At tree level they are massless and have no 
VEV. However, as we will show here, at one-loop a (tachyonic)
squared mass term can be generated inducing a spontaneous symmetry breaking. 
For simplicity, we will denote $H_M^0 =H_M$ and $H_M^{0*} ={\bar H}_M$ as 
they are the only components that will obtain a VEV. 
The generic mass terms for $H_5$ and $H_6$ are given by the 
coefficients of quadratic terms in the expansion of the effective 
Lagrangian around $H_5=H_6=0$,
\bea
\label{lag56}
-\L_{\rm mass} &=&
M_{5{\bar 5}}^2 |H_5|^2 + M_{6{\bar 6}}^2 |H_6|^2 + 
M_{5{\bar 6}}^2 H_5 {\bar H}_6 + M_{{\bar 5}6}^2 {\bar H}_5 H_6  \\
&=&M_{5{\bar 5}}^2 |H_5|^2 + M_{6{\bar 6}}^2 |H_6|^2 +  M_{+}^2 
(H_5 {\bar H}_6 
+ {\bar H}_5 H_6) + M_{-}^2 (H_5 {\bar H}_6 - {\bar H}_5 H_6)  \ ,
\nonumber
\eea
where we have defined:
\bea
M_{+}^2 = \frac{1}{2} (M_{5{\bar 6}}^2 + M_{{\bar 5}6}^2) \qquad  
M_{-}^2 = \frac{1}{2} (M_{5{\bar 6}}^2 - M_{{\bar 5}6}^2) 
\eea
Reality of the Lagrangian (\ref{lag56}) implies that $M_{\bar 5 5}$ and
$M_{\bar 6 6}$ are real and
$M_{\bar 5 6}=M_{\bar 6 5}^*$, so that $M_{+}^2$ is real, while
$M_{-}^2$ is purely imaginary.

However, since the fields $H_{5,\,6}$ do not have a well defined hypercharge,
we should write the Lagrangian for the standard model Higgs fields
$H_{1,\,2}$. Using, from Eq.~(\ref{Hdef}), 
$H_5= (\bar{H}_1 + H_2)/2 $ and $H_6=(\bar{H}_1 - H_2)/2i $, 
this part of the Lagrangian 
can be written as a function of $H_1$ and $H_2$
as:
\begin{equation}
-\L_{\rm mass}=\, 
m_{1}^2\ |H_1|^2 + m_{2}^2\ |H_2|^2 +  
\mu_{+}^2\ 
(H_1 H_2 + {\bar H}_1 \bar{H}_2) + \mu_{-}^2\ (H_1 H_2 - 
{\bar H}_1 \bar{H}_2) 
\label{lag12}
\end{equation}
with
\bea
m_1^2= \frac{1}{4} \left[M_{5{\bar 5}}^2 + M_{6{\bar 6}}^2\right] 
+ i\ M_{-}^2 && 
m_2^2= \frac{1}{4} \left[M_{5{\bar 5}}^2 + M_{6{\bar 6}}^2\right] 
- i \ M_{-}^2 \\
\mu_{+}^2 = \frac{1}{4} [M_{5{\bar 5}}^2 - M_{6{\bar 6}}^2]
 &&  
\mu_{-}^2 = i \ M_{+}^2 
\eea
where $m_{1}^2$, $m_{2}^2$ and $\mu_{+}^2$ are real while
$\mu_{-}^2$ is purely imaginary. The last two terms in (\ref{lag12}) can
be written in standard notation as $[m_3^2 \, H_1 H_2+h.c.]$ where 
$m_3^2= \mu_{+}^2+\mu_{-}^2$. If $\mu_{-}^2\neq 0$, $m_3^2$ is a complex
parameter and there is explicit $CP$-violation if the phase of $m_3^2$
cannot be absorbed into a redefinition of the Higgs fields. 

In general there can be one-loop generated quartic couplings, $\lambda_5$,
$\lambda_6$ and $\lambda_7$, in the effective potential that can prevent
such field redefinitions. They look like
\begin{equation}
-\L_{\rm quartic}=\frac{1}{2}\lambda_5 (H_1 H_2)^2
+(H_1\,H_2)\left[\lambda_6\, |H_1|^2+\lambda_7\, |H_2|^2\right]+h.c.
\label{quartic}
\end{equation}

However, in order to prevent tree-level flavor changing neutral currents
one usually enforces the discrete $\mathbb{Z}_2$ symmetry, $H_2\to -H_2$,
which is only softly violated by dimension-two operators, and prevents the
appearance of $\lambda_{6}$ and $\lambda_7$-terms, 
i.e. $\lambda_{6}=\lambda_{7}=0$~\cite{Hunter}.
In that case, the phase of $m_3^2$
cannot be absorbed into a redefinition of the Higgs fields provided that
$Im(\lambda_5^* m_3^4)\neq 0$, which signals $CP$-violation.

\subsection{Toroidal compactification} 

We will first compute the Higgs mass parameters $M_{5{\bar5}}^2$,
$M_{6{\bar6}}^2$, $M_{+}^2$ and $M_{-}^2$ induced at one-loop by the 
fermionic matter fields in the case of a compactification on a torus.
We denote by $G_{IJ}$ the torus metric as given in (\ref{metric}) 
and by $G^{IJ}$ its inverse. 
The interaction Lagrangian between the six-dimensional Weyl fermion 
$\Psi_\epsilon$, satisfying $\P_{\epsilon} \Psi_\epsilon = \Psi_\epsilon$ 
with $\epsilon = \pm$, with the Higgs fields:
\bea 
 - g \bar{\Psi}_{\epsilon} \Gamma^{M}   H_{M}  \Psi_{\epsilon}
\label{ferort}
\eea
induces, at one-loop, a quadratic term $H_{I}\bar H_{J}$
from the diagram of Fig.~3.
\begin{figure}[H]
\begin{center}
\SetScale{1.25}
\begin{picture}(120,60)(0,0)
\CArc(60,20)(20,0,360) 
\DashLine(0,20)(40,20)3 
\DashLine(80,20)(120,20)3 
\Text(0,35)[l]{$H_I$}
\Text(150,35)[r]{$\bar H_J$}
\Text(75,60)[c]{$\Psi_\epsilon$}
\end{picture}
\end{center}
\caption{One-loop diagram contributing to $M_{I{\bar J}}^2$.}
\end{figure}
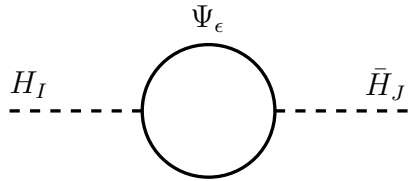

Calculation of the diagram of Fig.~3 yields the result:
\bea
M^2_{I\bar J }& =& 
 g^2   \sum_{{p}^5,\,{p}^6} 
\int \frac {d^4p}{(2\pi)^4} 
{\mbox{\rm Tr}}\left\{\Gamma^I\, \P_{\epsilon} \, 
\frac{1}{\slap{P}}\,  \, \Gamma^J  \, \P_{\epsilon}
 \, \frac{1}{\slap{P}}\,\right\} \nonumber \\
&=& - 4\, g^2\,R_{I}R_{J} \sum_{{p}^5,\,{p}^6} \int 
\frac {d^4p}{(2\pi)^4} \left\{ 
\frac{G^{IJ}}{p^2+ p_M G^{MN} p_N } -  
\frac{2 G^{IK} G^{JL} {p}_K {p}_L}
{(p^2+ p_M G^{MN} p_N )^2}
\right\} 
\label{massgen}
\eea
with
$\slap{P} = \Gamma_{\hat{\mu}}P^{\hat{\mu}}= \slap{p}+\Gamma_5 p^{5} 
+\Gamma_6 p^{6}$ 
where $p$ is the four-dimensional momentum. For simplicity of notations 
we use here $R_5\equiv R_1$ and  $R_6\equiv R_2$. 
The $R_{I}R_{J}$ factors arise 
because the normalization of the metric is such that the $p^I$ are integers.  
In the last equality of (\ref{massgen}) we have used the 
$\Gamma$-matrices property:
\bea
{\mbox{\rm Tr}}\left[\Gamma^I \P_{\epsilon}
\Gamma^{\hat{\mu}} \Gamma^J\P_{\epsilon} \Gamma^{\hat{\nu}}
\right]= \frac{1}{2} 
{\mbox{\rm Tr}}\left[\Gamma^I \Gamma^{\hat{\mu}}
\Gamma^J \Gamma^{\hat{\nu}}\right]
= 4 (g^{I\hat{\mu}}g^{J\hat{\nu}}+ g^{I\hat{\nu}}
g^{J\hat{\mu}} - g^{IJ}g^{\hat{\mu}\hat{\nu}})
\eea
where $g^{\hat{\mu}\hat{\nu}}$ has elements 
$\{ g^{\mu \nu}=\eta^{\mu \nu}, g^{\mu I}=0, g^{IJ}= G^{IJ}R_{I}R_{J}\}$ .

Note that the result of 
(\ref{massgen}) is independent of the six-dimensional 
chirality so we can choose $\epsilon =+$ without loss of generality.

To perform the integration in (\ref{massgen}) we use 
the Schwinger representation:
\bea
\frac{1}{(p^2+m^2)^n}=\frac{1}{\Gamma(n)}
\int_0^\infty dt\, \, t^{n-1} \, \, e^{- (p^2+m^2)t}
\eea
and make the change of variables $t = 1/l$. This gives:
\begin{equation}
M^2_{I{\bar J}}= -
\frac {4 g^2}{16 \pi^2}  R_{I}R_{J} \sum_{p_5,\, p_6} 
\int_0^\infty dl \, \, \, \,
\left(G^{IJ}-  \frac{2\, G^{IK} G^{JL} {p}_K {p}_L}{l}\right)\, \,  
e^{-( p_M  G^{MN}p_N )/2l}  
\end{equation}
As the momenta $p_5$ and $p_6$ take integer values, we can 
perform a Poisson ressumation which gives:
\bea
M^2_{I{\bar J}}= -
\frac {4 g^2 \pi}{8 } \sqrt{G} R_I R_J   
\sum_{{\tilde p}^5,\,{\tilde p}^6} 
\int_0^\infty dl  \, \, \, l^2   \, \, \, 
  {\tilde p}^I  {\tilde p}^J    \, \, \,  
e^{- \pi^2 ({\tilde p}^M  G_{MN} {\tilde p}^N )l}
\label{masaii}
\eea
where 
${\tilde p}_I$ are momenta on the dual lattice.  ${\tilde p}_I$ take 
integer values with our choice for the metric in (\ref{metric}).
Notice that the integrand dies exponentially when $l\to\infty$
except when ${\tilde p}^5={\tilde p}^6=0$. However
this term is zero in the summation (\ref{masaii}) because of the
prefactor ${\tilde p}^I  {\tilde p}^J$ and the integral is well
behaved for all values of ${\tilde p}^5,\,{\tilde p}^6$.
We finally make the change of variables 
$l'=\pi^2 ({\tilde p}^M  G_{MN} {\tilde p}^N )l$ 
and perform the integration on $l'$ to obtain:
\bea
M^2_{I\bar J}= - \frac{ N_F  g^2 }{4 \pi^5}  \sqrt{G}
\sum_{{\tilde p}^5,\,{\tilde p}^6} 
\frac{R_I R_J {\tilde p}^I  {\tilde p}^J}{\left[
{\tilde p}^M  G_{MN} {\tilde p}^N\right]^3}
\label{ftor}
\eea
with $N_F = 4$ being the number of 
degrees of freedom of a six-dimensional Weyl spinor.

This corresponds to the mass parameters:
\bea
M_{5{\bar 5}}^2&=& - 
\frac{ N_F  g^2 }{4 \pi^5} R_1 R_2 s  \sum_{n_1,\,n_2} 
\frac{n_1^2 R_1^2}{\left[  n_1^2 R_1^2 +n_2^2 R_2^2  
+2 c\, n_1 n_2 R_1  R_2\right]^3}, \nonumber \\
M_{6{\bar 6}}^2&=& - \frac{ N_F  g^2 }{4 \pi^5} R_1 R_2 s  \sum_{n_1,\,n_2} 
\frac{n_2^2 R_2^2}{\left[  n_1^2 R_1^2 +n_2^2 R_2^2  
+2 c\, n_1 n_2 R_1  R_2\right]^3},\nonumber \\
 M_{+}^2= M_{5{\bar 6}}^2 =M_{{\bar 5}6}^2&=& - 
\frac{ N_F  g^2 }{4 \pi^5} R_1 R_2 s \, \, \, \sum_{n_1,\,n_2} 
\frac{  n_1 n_2\, R_1\, R_2}{\left[  n_1^2 R_1^2 +n_2^2 R_2^2  +
2 c\, n_1 n_2\, R_1 R_2\right]^3}\ ,\nonumber \\
M_{-}^2 &=&0 \,
\eea
which implies that
\bea
m_{1}^2= m_{2}^2 &=& -
\frac{ N_F  g^2 }{8 \pi^5} R_1 R_2 s  \sum_{n_1,\,n_2} 
\frac{n_1^2 R_1^2 + n_2^2 R_2^2}{\left[  n_1^2 R_1^2 +n_2^2 R_2^2  
+2 c\, n_1 n_2 R_1  R_2\right]^3}, \nonumber \\
\mu_{-}^2&=& - i\frac{ N_F  g^2 }{4 \pi^5} R_1 R_2 s \, \, \, \sum_{n_1,\,n_2} 
\frac{ n_1 n_2\, R_1\, R_2}{\left[  n_1^2 R_1^2 +n_2^2 R_2^2  +
2 c\, n_1 n_2\, R_1 R_2\right]^3}\ ,\nonumber \\  
\mu_{+}^2 &=& 0 \,
\eea
It is easy to check that the CP-violating term  $\mu_{-}^2$ vanishes for 
$c=\cos{\theta}= 0$. This yields $m_3^2$ purely imaginary so that if the
quartic coupling $\lambda_5 (H_1\, H_2)^2+h.c.$ is generated with a real
coefficient then $Im(m_3^4\,\lambda_5^*)=0$ and there is no
$CP$-violation.

\subsection{Orbifold compactification}

Let us now turn to the orbifold case of our example. We will
carry the computation of the one-loop Higgs mass induced
by the fermions originating from a six-dimensional Weyl fermion $\Psi_+$:
\bea
\Psi_{(6)+}=\left[\begin{array}{l}\psi_L\\ 0
    \\ 0 \\ \Psi_{R }\end{array}
\right] \qquad {\rm with  \, }\qquad
\psi_{L} = \left(\begin{array}{l} {\tilde \phi_e}\\ \phi_e  \\ \phi_o 
\end{array} \right)
\qquad  {\rm and \, }\qquad 
\Psi_{R} = \left(\begin{array}{l} {\tilde {\bar{\chi}}_o}\\ {\bar{\chi}}_o
  \\ {\bar{\chi}}_e 
\end{array} \right)
\eea
The $\Gamma$-matrices $\Gamma^{\hat{\mu}}_\perp$ 
as written in (\ref{Gammas}) are given in an orthogonal
basis. In order to write the Yukawa interaction with the Higgs fields we need
to define the $\Gamma$-matrices in the basis associated
to $x^5,x^6$ and forming an angle $\theta$ (see Fig.~\ref{torus}):
\bea
\Gamma^5 &=& \Gamma^5_\perp - \frac{c}{s}\  \Gamma^6_\perp\nonumber\\
\Gamma^6 &=& \frac{1}{s}\  \Gamma^6_\perp  
\label{redefinicion}
\eea
which satisfy $\{ \Gamma^M,\Gamma^N \}= 2 g^{MN}$. The Yukawa 
interaction giving rise to masses for the components of the fermions 
in $\L$ can be obtained from the expansion of (\ref{ferort}):

\bea 
{\L_{\rm Yukawa}}&=&  -\, g \,\frac{1}{s}  \,  
\left( H_6 - e^{-i \theta} H_5\right)\, \, \,
 \bar{\phi_e} {\bar{\chi}}_e 
\, \, \, -\,  g \, \frac{1}{s} \,  
\left({\bar H}_6 - e^{i \theta} {\bar H}_5\right)\, \, \,
 {\chi}_e\phi_e \nonumber \\
&&- \, g \, \frac{1}{s} \,  
\left({\bar H}_6 - e^{- i \theta} {\bar H}_5\right)\, \, \, 
\bar{\phi_o} {\bar{\chi}}_o 
\, \, \,- \, g \, \frac{1}{s} \,  
\left(H_6 -  e^{i \theta} H_5\right)\, \, \,
{\chi}_o \phi_o
\label{eEH1}
\eea

As we are performing the computation in the symmetric phase, 
i.e. an expansion around $H_M=0$, we can use the free fields 
Kaluza-Klein decomposition of the fermion fields.
The $\mathbb{Z}_2$-even states $\phi_e$ and ${\bar{\chi}}_e$ have the 
following decomposition:
\bea
\phi_e &=& \frac{1}{\pi \sqrt {R_1 R_2}} \sum_{n_1 = - \infty}^{\infty} 
\sum_{n_2 = - \infty}^{\infty}  \cos\left( {\frac {n_1 x^5}{R_1}}+
{\frac {n_2 x^6}{R_2}}\right)
\phi^{(n_1,n_2)}_e\nonumber \\
{\bar{\chi}}_e &=& \frac{i}{\pi \sqrt {R_1 R_2}} 
\sum_{n_1 = - \infty}^{\infty} 
\sum_{n_2 = - \infty}^{\infty} 
\cos\left( {\frac {n_1 x^5}{R_1}}+
{\frac {n_2 x^6}{R_2}}\right)
{\bar{\chi}}^{(n_1,n_2)}_e
\eea
while for the $\mathbb{Z}_2$-odd states $\phi_o$ and ${\bar{\chi}}_o$ we have,
\bea
\phi_o  &=& \frac{1}{\pi \sqrt {R_1 R_2}} \sum_{n_1 = - \infty}^{\infty} 
\sum_{n_2 = - \infty}^{\infty} \sin\left( {\frac {n_1 x^5}{R_1}}+
{\frac {n_2 x^6}{R_2}}\right)
\phi^{(n_1,n_2)}_o\nonumber \\
{\bar{\chi}}_o&=& \frac{i}{\pi \sqrt {R_1 R_2}} \sum_{n_1 = - \infty}^{\infty} 
\sum_{n_2 = - \infty}^{\infty} 
\sin\left( {\frac {n_1 x^5}{R_1}}+
{\frac {n_2 x^6}{R_2}}\right)
{\bar{\chi} }^{(n_1,n_2)}_o
\label{fourier}
\eea
where the transformation properties under the orbifold group action imply: 
\bea
\phi^{(-n_1,-n_2)}_e= \phi^{(n_1,n_2)}_e && \phi^{(-n_1,-n_2)}_o= 
- \phi^{(n_1,n_2)}_o 
\nonumber\\
{\bar{\chi}}^{(-n_1,-n_2)}_e =  {\bar{\chi}}^{(n_1,n_2)}_e &&
{\bar{\chi} }^{(-n_1,-n_2)}_o = - {\bar{\chi} }^{(n_1,n_2)}_o
\label{iden}
\eea

The Yukawa couplings of $H_5$ and $H_6$ are given by:
\bea
\int d^4x \int_{0}^{\pi R_1} dx^5 \int_{0}^{\pi R_2} dx^6 
\left[  g H_5 \left(e^{-i \theta} \bar{\phi_e} {\bar{\chi}}_e + 
e^{i \theta}{\chi}_e\phi_e\right)
- g H_6 \left( \bar{\phi_e} {\bar{\chi}}_e
 + {\chi}_e\phi_e\right)
\right] \nonumber \\
\label{HeE}
\eea 
while those of ${\bar H}_5$ and ${\bar H}_6$ can be
obtained by complex conjugation. Using the KK-decomposition of
(\ref{fourier}) in (\ref{eEH1}) we obtain the coupling of $H_6$ to
the KK-excitations of the fermion: 
\begin{equation}
{\L_{Y6}}=- \sum_{n_1,\, n_2}\,\!\!\!^\prime
\ i g  H_6 \left({\bar{\phi}
}^{(n_1,n_2)}_e  {\bar{\chi} }^{(n_1,n_2)}_e + {\bar{\phi}
}^{(n_1,n_2)}_o  {\bar{\chi} }^{(n_1,n_2)}_o\right)+ h.c.  
\label{lag}
\end{equation}
where the summation $\sum\,\!\!^\prime$ is defined as,
\begin{equation}
\label{definicion}
\sum_{n_1,\, n_2}\,\!\!\!^\prime\
\ f(n_1,\,n_2)=
\sum_{n_1=1}^{\infty} \ \sum_{n_2=-\infty}^\infty\  f(n_1,\,n_2)
+\sum_{n_2=0}^\infty\  f(0,\,n_2)
\end{equation}

Notice that there is no overcounting of states in (\ref{lag}) since,
from (\ref{iden}), $\phi_o^{(0,0)}=\chi_o^{(0,0)}\equiv 0$. It is also
important to note that even for this orbifold case the summation can
be made on a full tower of KK-excitations  $n_1, n_2 \in [- \infty,
\infty ] $. This can be made explicit by defining 
$\phi^{(n_1,n_2)}$ and $\chi^{(n_1,n_2)}$ through: 
\bea
\phi^{(n_1,n_2)}=  \left\{ \begin{array}{ll} \phi^{(n_1,n_2)}_e & {\rm
for} \ n_1>0 \\ \phi^{(n_1,n_2)}_e & {\rm for} \ n_1=0, \,
n_2 >0 \\ \phi^{(0,0)}_e & {\rm for} \ n_1=n_2 =0 \\
\phi^{(n_1,n_2)}_o & {\rm for} \ n_1=0, \,  n_2 <0 \\
\phi^{(n_1,n_2)}_o  & {\rm for}\ n_1<0 \end{array} \right. ,
\ \chi^{(n_1,n_2)}= \left\{ \begin{array}{ll} \chi^{(n_1,n_2)}_e &
{\rm for} \ n_1>0 \\ \chi^{(n_1,n_2)}_e & {\rm for} \ n_1=0,
\,  n_2 >0 \\ \chi^{(0,0)}_e & {\rm for} \ n_1=n_2 =0 \\
\chi^{(n_1,n_2)}_o & {\rm for} \ n_1=0, \,  n_2 <0 \\
\chi^{(n_1,n_2)}_o  & {\rm for}\ n_1<0 \end{array}
\right. \nonumber \\ 
\eea 

The interaction Lagrangian between $H_6$ and
the fermions takes then the form:
\begin{equation} 
\L_{Y6}=
- \sum_{n_1 = - \infty}^{\infty}  \sum_{n_2 = - \infty}^{\infty} i g
\left[H_6 {\bar{\phi} }^{(n_1,n_2)}  {\bar{\chi} }^{(n_1,n_2)} + {\bar
H}_6 {{\phi} }^{(n_1,n_2)} {{\chi} }^{(n_1,n_2)}\right]
\label{ly6}
\end{equation}
The diagonal
one-loop induced mass term $M_{6 {\bar 6}}^2 |H_6|^2$ is then automatically
finite, as it is due to a whole tower of KK states. In fact the
contribution of 
a full tower can be computed directly from the toroidal case (\ref{ftor})
with the replacement $N_F \rightarrow N_F({\rm orbifold})
= \frac{1}{2}N_F({\rm torus})$. 

In the same way, the Yukawa coupling of $H_5$ with fermions of 
positive six-dimensional chirality is given by:
\begin{equation}
{\L_{Y5}}= 
\sum_{n_1,\, n_2}\,\!\!\!^\prime\
 g  H_5 \left[ i e^{- i \theta}
 {\bar{\phi} }^{(n_1,n_2)}_e 
{\bar{\chi} }^{(n_1,n_2)}_e +i e^{i \theta} {\bar{\phi} }^{(n_1,n_2)}_o 
{\bar{\chi} }^{(n_1,n_2)}_o\right]+ h.c.
\label{ly5}
\end{equation}
When computing the one-loop diagrams of Fig.~3 contributing to
$M_{5{\bar 5}}^2$, the product of phases $e^{\pm i \theta}$ at
the two vertices cancel to each other 
and the result also corresponds to the contribution of a whole 
tower of states. It can be obtained from $M_{6{\bar 6}}^2$
through the exchange of $R_1$ with $R_2$.
In fact, it is possible to write the Lagrangian (\ref{ly5}) as 
a Yukawa coupling interaction of $H_5$ with a complete tower of 
KK excitations, as it was done in (\ref{ly6}) for $H_6$, 
by making phase rotations on the fermions. 
However, for $\theta \neq \frac{\pi}{2}$ 
one cannot write simultaneously both (\ref{ly5}) and (\ref{ly6})
as interactions with a whole tower. In fact, the phase $e^{\pm i \theta}$
comes from the metric of the torus and for $\theta \neq \frac{\pi}{2}$ it 
is at the origin of the appearance of a $M_{-}^2$ term as we will see now.

The one-loop induced mixing terms between $H_5$ and $H_6$ can 
also be computed in a straightforward manner as the sum of
the contribution of even states and that from odd states
propagating in the loop of Fig.~3.
The result of the one-loop mixing diagrams can be written formally as:
\bea
M_{5{\bar 6}}^2 &=& e^{-i\theta}\left(\begin{array}{l} 
{\rm contributions\, \, of} \\
{\rm \, \, even\, \, states} \end{array} \right) + e^{i\theta} 
\left(\begin{array}{l} {\rm contribution \, \, of} \\
{\rm \, \,odd \, \, states} \end{array} \right)\\
M_{{\bar 5}6}^2 &=& \, \, 
e^{i\theta}\left(\begin{array}{l} {\rm contribution\, \, of} \\
{\rm \, \, even \, \, states} \end{array} \right) + e^{-i\theta} 
\left(\begin{array}{l} {\rm contribution \, \, of} \\
{\rm \, \, odd \, \, states} \end{array} \right)
\label{eme56}
\eea 
which implies: 
\bea
M_{+}^2 = c \left[ \left(\begin{array}{l} {\rm contributions\, \, of} \\
{\rm \, \, even\, \, states} \end{array} \right) +
\left(\begin{array}{l} {\rm contribution \, \, of} \\
{\rm \, \,odd \, \, states} \end{array} \right) 
\right] 
\label{problem1}
\eea
\bea
M_{-}^2 = - i\, \,  s  \left[ \left(\begin{array}{l} {\rm contributions\, \, of} \\
{\rm \, \, even\, \, states} \end{array} \right) -
\left(\begin{array}{l} {\rm contribution \, \, of} \\
{\rm \, \,odd \, \, states} \end{array} \right) 
\right] 
\label{emeplus}
\eea

In (\ref{problem1}) 
the sum  of contributions from even and odd states
reproduces the one from a whole tower of states. The overall 
$c$ is necessary to reproduce the metric factor in the product of the two
momenta $p_5$ and $p_6$ (see the double product in (\ref{massgen})). 
We then reproduce for $M_{+}^2$ the result of the torus with, again, 
$N_F \rightarrow N_F({\rm orbifold}) = \frac{1}{2}N_F({\rm torus})$.

Next, we consider the mass parameter $M_{-}^2$. Due to the relative sign in  
(\ref{emeplus}), all the 
contributions of even and odd massive KK-states cancel to each other
and only the divergent contribution of the massless mode remains!
While each tower of KK excitations of the massless fermions contributes 
with a divergent result, the sum of all of these contributions is finite 
and, in our case, it vanishes. Indeed, the cancellation of irreducible 
non-abelian anomalies in six-dimensions (see next section) 
requires the fermions
to arise from six-dimensional Weyl spinors that can be paired with opposite  
six-dimensional chiralities. While the Higgs field interaction with 
the positive chirality fermions is given by:
\bea
{\L^+_{\rm Yukawa}}= -\, g \,\frac{1}{s}  \,  
\left( H_6 - e^{-i \theta} H_5\right)\, \, \,
 \bar{\phi_e} {\bar{\chi}}_e 
- \, g \, \frac{1}{s} \,  \left(H_6 -  e^{i \theta} H_5\right)\, \, 
{\chi}_o \phi_o\, \, + h.c 
\label{ferp+}
\eea
the ones with negative chirality fermion interaction is given by
\bea
{\L^-_{\rm Yukawa}}= -\, g \,\frac{1}{s}  \,  
\left( H_6 - e^{i \theta} H_5\right)\, \, \,
 \bar{\phi_e} {\bar{\chi}}_e 
- \, g \, \frac{1}{s} \,  \left(H_6 -  e^{-i \theta} H_5\right)\, \, 
{\chi}_o \phi_o\, \, + h.c 
\label{ferp-}
\eea
The contribution of fermions originating from
six-dimensional spinors with negative chirality can be obtained from  
the one due to spinors with positive chirality through the 
exchange of $e^{i \theta} \rightarrow e^{-i \theta}$. For each pair of such 
fermions the contribution to $M_{-}^2$ cancels. In our model of section 4,
we have the leptons $L$ whose contribution is canceled by that
of $Q_3$, and the quarks $Q_2$ that cancel the contributions from $Q_1$.
The sum of the contributions of all fermions leads then to $M_{-}^2 =0$.

Our results for the mass parameters are then:
\bea
N_F &\rightarrow& N_F({\rm orbifold}) = \frac{1}{2}N_F({\rm torus}) \nonumber\\
M_{5{\bar 5}}^2&=& - 
\frac{ N_F  g^2 }{4 \pi^5} R_1 R_2 s  \sum_{n_1,\,n_2} 
\frac{n_1^2 R_1^2}{\left[  n_1^2 R_1^2 +n_2^2 R_2^2  
+2 c\, n_1 n_2 R_1  R_2\right]^3}, \nonumber \\
M_{6{\bar 6}}^2&=& - \frac{ N_F  g^2 }{4 \pi^5} R_1 R_2 s  \sum_{n_1,\,n_2} 
\frac{n_2^2 R_2^2}{\left[  n_1^2 R_1^2 +n_2^2 R_2^2  
+2 c\, n_1 n_2 R_1  R_2\right]^3},\nonumber \\
 M_{+}^2= M_{5{\bar 6}}^2 =M_{{\bar 5}6}^2&=& - 
\frac{ N_F  g^2 }{4 \pi^5} R_1 R_2 s \, \, \, \sum_{n_1,\,n_2} 
\frac{  n_1 n_2\, R_1\, R_2}{\left[  n_1^2 R_1^2 +n_2^2 R_2^2  +
2 c\, n_1 n_2\, R_1 R_2\right]^3}\ ,\nonumber \\
M_{-}^2 &=&0 \,
\eea
which implies that
\bea
m_{1}^2= m_{2}^2 &=& -
\frac{ N_F  g^2 }{8 \pi^5} R_1 R_2 s  \sum_{n_1,\,n_2} 
\frac{n_1^2 R_1^2 + n_2^2 R_2^2}{\left[  n_1^2 R_1^2 +n_2^2 R_2^2  
+2 c\, n_1 n_2 R_1  R_2\right]^3}, \nonumber \\
\mu_{-}^2&=& - i\frac{ N_F  g^2 }{4 \pi^5} R_1 R_2 s \, \, \, \sum_{n_1,\,n_2} 
\frac{ n_1 n_2\, R_1\, R_2}{\left[  n_1^2 R_1^2 +n_2^2 R_2^2  +
2 c\, n_1 n_2\, R_1 R_2\right]^3}\ ,\nonumber \\  
\mu_{+}^2 &=& 0 \,
\label{radmasas}
\eea

It is important to note that results of the diagrammatic 
one-loop computation exactly reproduce the results of the expansion of 
the one-loop effective potential in section 3 upon identification: 
$H_5={\bar H_5}=a_1/g R_1$ and $H_6={\bar H_6}=a_2/g R_2$.

The geometrical origin of the mixing terms between 
$H_5$ and $H_6$ can be understood easily. Upon toroidal 
compactification, the six-dimensional Lorentz invariance is broken and 
one is left with translation invariance along the two internal dimensions. 
This symmetry is enough to forbid transitions between the components 
$A'_5$ and $A'_6$ of the internal gauge field in an orthogonal basis
and thus forbids any mass term of the form $A'_5 A'_6$. 
However, due to the presence 
of the angle $\theta \neq \frac {\pi}{2}$ the new fields $A_6$ have a 
component $c A_6$ along the fifth dimension, 
which implies transitions amplitudes
between $A_6$ and $ A_5$, or equivalently between $H_5$ and $H_6$, 
$c =\cos{\theta}\neq 0$. On the other hand, the mass terms 
$M_{-}^2$ ($\mu_{-}^2$) correspond to transitions between 
elements of the orthogonal
basis which do not receive contributions from the bulk fields
$M_{-}^2=\mu_{+}^2 = 0$.
However, upon orbifolding of the torus, 
the translation invariance is broken at 
the boundaries and then terms mixing $A'_5$ and $A'_6$
could a priori appear localized on the fixed points through higher loops
involving localized states.

\section{Higgs potential from $U(1)$ anomaly cancellation}

It is easy to show that the six-dimensional model with the spectrum given in
(\ref{repr}) is free from irreducible 
anomalies~\footnote{As the non-abelian factors in our model are 
$SU(3)$'s there is no irreducible $\tr F^4$. However, there is the 
possibility to have terms of the form $\tr \Q_i F^3$ }. Indeed the associated
anomaly polynomial which describes all mixed non-abelian and
gravitational anomalies is factorizable and it is given by:
\bea
A_{6D}= \tr{F_c^2}\, \tr{F_w^2}
\label{6dano}
\eea
where $F_c$ and $F_w$ are the strength fields of the $SU(3)_c$ and 
$SU(3)_w$ gauge fields respectively. The reducible anomaly 
(\ref{6dano}) can be canceled by a generalized 
Green-Schwarz mechanism \cite{GS}.

The compactification to four dimensions on the $T^2/\mathbb{Z}_2$ orbifold 
considered in section 4 does not produce any non-abelian anomalies. 
The chiral spectrum obtained through the $\mathbb{Z}_2$ 
projection leads however to anomalies for the $U(1)$ factors. The 
mixed anomalies of the three $U(1)$'s with non-abelian factors are given 
by the matrix $[\A_{ai}]$:
\bea
[\A_{\alpha i}]=\left(\begin{array}{ccc}\,-{6}&-\frac{3}{2}&0\\
-{{3}}&-3&\frac{9}{2}\,\end{array} \right)  
\eea
where  $\A_{\alpha i} =\tr (\Q_i t^2_\alpha)$ with
$t_1$ and $t_2$  the generators of $SU(3)_c$ and $SU(2)_w$ respectively. 
It is easy to check that one 
$U(1)$ combination, corresponding to the hypercharge $U(1)_Y$:
\bea 
\Q_Y =\frac{1}{3}\left(\frac{1}{2}\Q_1   - 2 \Q_2 - \Q_3\right)
\eea
obtained in (\ref{hypercharge}), is anomaly free while 
the two other orthogonal combinations of $U(1)$ factors
\bea
\Q'&=&\frac{1}{\sqrt{30}}(2 \Q_1 + \Q_3)
 \nonumber \\
\Q''&=& \frac{1}{\sqrt{30}}(\Q_2 - 2 \Q_3)
\eea
are anomalous.
These anomalies can generally be canceled in two possible ways:
(i) By the appearance of extra matter localized at the orbifold fixed 
points with the appropriate quantum numbers to cancel the anomalies; 
(ii) By a generalized Green-Schwarz mechanism. 

Although, for simplicity, we will only consider below the second
possibility, the former one could also be easily realized. For instance, 
if the model originates from D5-(anti)branes in type IIB orientifolds,
then D3-(anti)branes should also be introduced because of the
$\mathbb{Z}_2$ orbifold. 
Open strings with one end on these branes and the other
on the $U(3)\times U(3)$ D5-branes would give rise to extra matter
fields needed to cancel the $U(1)$ anomalies. As stated in section 2, a
sufficient condition to keep the one-loop Higgs mass finite is that these
localized states must appear in supermultiplets. In this case, the
results for the one-loop Higgs potential obtained above remain unchanged.

A way to avoid the appearance of extra branes and matter, is
by making the $\mathbb{Z}_2$ orbifold freely acting, 
combining for instance its
action with a shift by half a compactification lattice vector. Our
computation can be easily generalized for this case.

The generalized Green-Schwarz mechanism to cancel the above anomalies 
rests on the observation that if the model is obtained  using a string 
construction there should exist two (Ramond-Ramond) axion 
fields $a'$ and $a'' $ which 
transform non-trivially under the $U(1)'$ and $U(1)''$
gauge transformations in order to cancel the 
anomalies~\cite{DSW}. The couplings of these fields to the 
corresponding gauge fields, $B'_\mu$ and $B''_\mu$ is given by
the Lagrangian:
\begin{align} 
\label{kano}
\L &= -\frac{1}{2}(\partial_\mu a' + \lambda M_s B'_\mu)^2
-\frac{1}{2}\sum_k(\partial_\mu a'' + \lambda M_s B''_\mu)^2 \\
 & -\frac{1}{32\pi^2} \ \frac{a'}{ \lambda M_s}\ \sum_a k'_a \ 
F^{(a)}_{\mu\nu}\tilde F^{(a)\mu\nu} 
-\frac{1}{32\pi^2}\ \frac{a''}{ \lambda M_s}\ \sum_a \
k''_a  \
F^{(a)}_{\mu\nu}\tilde F^{(a)\mu\nu} 
\nonumber
\end{align}
where $\lambda$ is a parameter that depends on the string model and
\bea
k'_1= \tr \left(\Q' t^2_1\right)\qquad  k'_2=\tr \left(\Q' t^2_2\right) 
\qquad k''_1 = \tr \left(\Q'' t^2_1\right)
\qquad k''_2 = \tr \left(\Q'' t^2_2\right)
\eea
In order to cancel the phase from the fermionic determinant, 
the axions $a'$ and $a''$ need to transform as:
\bea 
\label{phi_transf}
\delta B'_\mu=\partial_{\mu} \Lambda', \quad\ 
\delta a' = g \lambda  M_s \Lambda'\qquad  {\rm and } \qquad \delta B''_\mu=\partial_{\mu} \Lambda'',
 \quad\ \delta a'' = g \lambda M_s \Lambda''
\label{transf}
\eea 

In the analysis of the modifications of the Higgs potential 
due to the use of a Green-Schwarz mechanism for cancelling the anomalies, 
it is necessary to make use of a basis for the  $U(1)$ charges. 
A possible choice 
is to use $U(1)_{Y}$, $U(1)_{\Q'}$ and $U(1)_{\Q''}$ such that the anomaly 
free combination 
$\Q_Y$ is made manifest. Instead, it is more convenient for our discussion 
to use $U(1)_{1}$ ($\Q_1$) since the 
Higgs fields do not carry charges under the other two
$U(1)$'s.
In order to obtain the modification to the Higgs potential we 
assume that the theory is obtained as a truncation of a 
supersymmetric (``super-parent'') 
theory projecting away the $R$-symmetry odd gauginos, sleptons, 
squarks and Higgsinos 
while keeping the $R$-parity even states: gauge bosons, matter 
fermions and 
Higgs scalars. The tree-level Lagrangian can then be obtained by putting 
to zero the $R$-odd states. This way of describing the model
as a truncation allows to obtain the 
modification for tree level Higgs potential easily. 
Indeed, this potential given in (\ref{vtree}) arises in the super-parent model
as the $D$-term potential. Its modification due to the cancellation of 
anomalies through the Green-Schwarz mechanism is well known, and given by:
\bea
&& \frac{g^2}{8} \left( |H_1^0|^2 - |H_2^0|^2 \right)^2 + \frac{3 \, g^2}{8} 
\left( |H_1^0|^2 - |H_2^0|^2 \right)^2 \nonumber \\ 
\rightarrow && \frac{g^2}{8} \left( |H_1^0|^2 - |H_2^0|^2 \right)^2 
+ \frac{3 \, g^2}{8} 
\left( |H_1^0|^2 - |H_2^0|^2 + \xi \frac{\varphi}{\sqrt{G}} \right)^2 
\label{varpi}
\eea
where $\xi$  is proportional to $\lambda$ and $\varphi$ is a scalar modulus
blowing up the orbifold singularities; it is complexified with the axion $a'$. 
The first term in (\ref{varpi}) arises from the $D$ term of the $U(1)$ in the 
Cartan of $SU(2)_w$, which is free of anomalies, while the second one arises 
from the $D$ term of $U(1)_1$, with anomalies canceled by the Green-Schwarz 
mechanism. The presence of $\frac{1}{\sqrt{G}}$ in (\ref{varpi}) is due to 
the absence of the $U(1)$ anomaly in the decompactification limit. 

The leading terms in the expansion of the scalar potential 
in powers of $H_1$ and $H_2$ are then given by:
\bea 
V_{\rm total}= V_c(\varphi, G_{IJ})+ V_0(H_1,H_2,\varphi, G_{IJ})
+\Delta V_1
\label{total}
\eea
where $V_0$ is the tree-level potential including the $U(1)$-anomaly,
Eq.~(\ref{varpi}), and $\Delta V_1$ is the one-loop effective potential from
bulk field loops, computed in previous sections.
We do not minimize with respect to $\varphi$, the internal metric 
$G_{IJ}$ and other moduli, as their corresponding effective potential  
$V_c$ is unknown. Instead, we consider these moduli as given parameters of
the theory and carry the minimization only with respect to $H_1$ and $H_2$.

We will start by analyzing the structure of $V_0$ as a function of the four
real fields $A_1=A_5^{(6)}$, $A_2=A_6^{(6)}$, $B_1=A_5^{(7)}$ and
$B_2=A_6^{(7)}$, in terms of which the neutral components of Higgs doublets
are defined as,
\begin{eqnarray}
H_5&=A_1+i\, B_1,\quad &H_1=\left[A_1-B_2-i(A_2+B_1) \right]/2 \nonumber\\
 H_6&=A_2+i\, B_2,\quad &H_2=\left[A_1+B_2-i(A_2-B_1) \right]/2 
\label{h56}
\end{eqnarray}
The potential $V_0$ reads,
\begin{equation}
V_0=\alpha\,(B_1\, A_2-A_1\, B_2)^2+\beta\,
(B_1\, A_2-A_1\, B_2+\xi \frac{\varphi}{\sqrt{G}} )^2
\label{vsub0}
\end{equation}
where $\alpha=g^2/8$ and $\beta={g'}^2/8$.
Notice that the VEVs of the $A_I$-fields are, after a trivial 
rescaling by $g R_I$, the Wilson line background we introduced in section 3.
In that case, i.e. for $B_I\equiv 0$, the potential $V_0$ is just a constant
provided by the anomaly.

Minimization with respect to $A_I$ and $B_I$ yields the condition
for the corresponding VEVs, $\langle A_I \rangle =a_I$,
$\langle B_I \rangle =b_I$
\begin{equation}
\label{minimo}
a_1\, b_2-b_1\, a_2=\kappa^2,\quad
\kappa^2=\frac{{g'}^2}{g^2+{g'}^2}\ \xi \frac{\varphi}{\sqrt{G}}
\end{equation}
and the squared mass matrix at the minimum is given by:
\begin{equation}
\label{masamat}
\mathcal{M}^2=2\,(\alpha+\beta)\left[
\begin{array}{cccc}
b_2^2 & -b_1\, b_2 & -b_2\, a_2 & a_1\, b_2\\
-b_1\,b_2 & b_1^2 & b_1\, a_2 & -b_1\, a_1\\
-b_2\, a_2 & b_1\, a_2 & a_2^2 & -a_1\, a_2\\
a_1\, b_2 & -b_1\, a_1 & -a_1\, a_2 & a_1^2
\end{array}
\right]
\end{equation}
where the VEVs $a_I$ and $b_I$ are subject to the condition (\ref{minimo}).
The matrix (\ref{masamat}) has one non-vanishing mass eigenvalue, given by
\begin {equation}
\label{masan0}
M^2=\frac{g^2+{g'}^2}{2}\, v^2\ ,
\end{equation}
where $v^2\equiv |H_1|^2+|H_2|^2$, and three massless eigenvalues 
corresponding to three flat directions of the potential $V_0$.
The mass eigenstates are
\begin{align}
\widetilde A_1=&-\frac{a_1}{b_2}\, A_1+B_2\label{flat1}\\
\widetilde A_2=&\ \frac{a_2}{b_2}\, A_1+B_1\label{flat2}\\
\widetilde B_1=&\ \frac{b_1}{b_2}\, A_1+A_2\label{flat3}\\
\widetilde B_2=&\ \frac{b_2}{a_1}\, A_1-
\frac{b_1}{a_1}\, A_2
-\frac{a_2}{a_1}\, B_1
+B_2
\label{nonflat}
\end{align}
where $\widetilde A_1$, $\widetilde A_2$ and $\widetilde B_1$ are the flat
directions of $V_0$.

If we define the $\beta$-angle as $\tan\beta=|H_2|/|H_1|$, we can use the
equation of minimum (\ref{minimo}) to write,
\begin{equation}
\label{tbeta}
\tan\beta=\sqrt{\frac{v^2+\kappa^2}{v^2-\kappa^2}}\ .
\end{equation}
In particular, in the absence of anomaly $\xi=0$ and $\tan\beta=1$.

Of course, the latter result is based on the tree-level minimization condition 
(\ref{minimo}) and radiative corrections, corresponding to the introduction
of the potential $\Delta V_1$, can provide small corrections to it. In 
particular we can introduce just the radiative mass terms of (\ref{radmasas})
in the potential $\Delta V_1$, neglect the one-loop generated quartic
couplings compared to the tree-level quartic potential, and assume 
that the determination of 
$\tan\beta$ from (\ref{minimo}) is a good enough approximation. The effective
potential, written as a function of $H_1$ and $H_2$ contains now a 
term as $\mu_-^2\, H_1\, H_2+h.c.$ where $\mu_-^2$ is purely imaginary as given
in (\ref{radmasas}). In fact, if we define $\mu_-^2\equiv i m_3^2$ we can
absorb the phase $e^{i\pi/2}$ into the Higgs product $H_1\,H_2$ and, since
$\lambda_6=\lambda_7=0$ in (\ref{total}), 
our approximated potential does not contain any 
explicit $CP$-violation. Using now the $SU(2)$ gauge invariance in order to
rotate one of the Higgs field VEVs on its real part, the remaining degrees of
freedom are $|H_1|$, $|H_2|$ and a phase, whose VEV would signal spontaneous
$CP$ breaking. However, since $\lambda_5=\lambda_6=\lambda_7=0$ 
in (\ref{total}), it is easy
to see that the dynamical phase is driven to zero. Minimization conditions
imply now,
\begin{equation}
\label{minimo1}
|H_1|^2-|H_2|^2+\kappa^2=\frac{4}{g^2+g^{\prime\,2}}\, 
\cot 2\beta\ m_3^2
\end{equation}
which generalizes Eq.~(\ref{minimo}). In fact, in the absence of $U(1)$
anomalies, for $\xi=0$, Eq.~(\ref{minimo1}) is only consistent for
$\tan\beta=1$, in agreement with Eq.~(\ref{tbeta}). In a sense 
Eq.~(\ref{minimo}) can be seen as the limit of (\ref{minimo1}) when 
$R_I\to\infty$. However, for finite radii Eq.~(\ref{minimo1}) can be used,
if $\tan\beta$ is approximately fixed by (\ref{tbeta}), to relate the
compactification radii and the physical VEV of the Higgs fields
$v^2=|H_1|^2+|H_2|^2$. We will do this for the case of equal radii, 
$R_1=R_2=R$, and an arbitrary torus angle $c\equiv \cos\theta$. Using the 
relations (\ref{minimo1}) and (\ref{radmasas}) we can write the 
compactification radius as a function of $c$ and the other parameters of the
theory, as $v^2$, $\kappa^2$ and $\tan\beta$,
\begin{equation}
\label{radio}
\frac{1}{R}=f(c)\ \sqrt{(10/N_F)\left(v^2\sin 2\beta
+\kappa^2\tan 2\beta\right)}\ .
\end{equation}

We have arbitrarily normalized $N_F$ to 10 and the function 
$f(c)$, that can be easily 
obtained from (\ref{radmasas}), has been plotted in Fig.~4, where we have
chosen $g^2=g^{\prime\,2}$.
\begin{figure}[H]
\centering
\epsfig{file=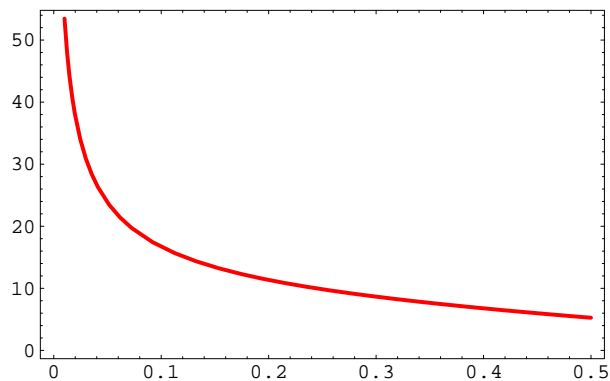,width=.5\linewidth}
\caption{The function $f(c)$ in (\ref{radio}).}
\label{f}
\end{figure}

From Fig.~4 we can see that, depending on the value of $c$, there is an
enhancement factor for the compactification scale $1/R$ with respect to the
weak scale $v$. This enhancement factor goes to $\infty$ when 
$c\to 0$~\footnote{For the case $c=0$, $m_3^2=0$ and Eq.~(\ref{minimo1})
goes back to (\ref{minimo}), for which the relation between $1/R$
and the weak scale is lost.}, which shows that we can obtain compactification
scales larger than the weak scale $1/R\gg v$ for a range of torus angles.
As we have seen in section 3 this enhancement factor 
disappears for pure Wilson lines since in that case the background field is
along the direction $|H_1|=|H_2|$ and all quartic (non-radiative) contributions
to the effective potential vanish.

\section{Discussion}
In this work we have studied the possibility that the standard model
Higgs boson would be identified with the component of a gauge field
along a compact extra dimension. The nice feature in such scenario is that
the Higgs mass is expected to be free of one-loop
quadratic divergences. Such divergences would introduce
a mass to the Higgs field that would not vanish
in the decompactification limit, and they are thus forbidden by the higher
dimensional local gauge symmetry.

Although higher dimensional gauge theories are non-renormalizable,
we have shown that for toroidal compactifications the full one-loop
potential of the Higgs field can be explicitly computed without any
reference to the underlying fundamental theory. As these toroidal
compactifications do not lead to a chiral spectrum, it is
necessary to introduce more complicated internal spaces. We considered
here compactification on an orbifold obtained from the torus by gauging a
discrete $\mathbb{Z}_2$ symmetry of the model.
The finiteness of the one-loop Higgs mass
is no more guaranteed in this case because of the presence of subspaces
fixed under the orbifold where the local higher
dimensional gauge symmetry is not conserved. Indeed,
in existing string examples, one often obtains massless states
localized at the orbifold fixed points in representations of the four
dimensional (but not the higher dimensional) gauge group.
We have shown that the one-loop result remains insensitive to the UV
theory if the localized matter appears degenerate between fermions and
bosons, forming $N=1$ supersymmetric multiplets.
Such a situation appears for instance in the class of non-supersymmetric
string models that were studied in~\cite{ABQ1}.

For such orbifold models we have 
computed the one-loop Higgs mass, both from the analysis of the effective 
potential and from a diagrammatic one-loop computation, and shown to
agree. The former method allows to compute the full one-loop effective 
potential dependence on tree-level flat directions. 
Instead, in the second (diagrammatic) approach 
we are able to compute the quadratic part for all scalar fields, 
however only as an expansion around 
the symmetric phase where the VEVs vanish.

In a fully realistic model the fermion flavor should be incorporated
from the fundamental theory. In fact, as the Higgs is identified with
an internal component of a gauge field, all tree level Yukawa
couplings are given by the gauge coupling and all particles
interacting with the Higgs field participate equally in generating its
mass. This is to be contrasted with the usual case where the one-loop
Higgs mass is dominated by the top quark due to the hierarchy of
Yukawa couplings. A possible approach would be to identify the two light
generations with (supersymmetric) boundary states with no tree level
Yukawa couplings. In this work we did not attempt to address the
problem of hierarchy of fermion masses. Instead, we tried to build a
simple model from compactifications on orbifold in order to
illustrate the main features of the scenario. We constructed a
model where the massless representations are exactly the ones of the
standard model, with two Higgs doublets originating from the internal
components of a gauge field. It was obtained as a compactification of
a six-dimensional model with gauge group $U(3)\times U(3)$ on a
$T^2/\mathbb{Z}_2$ orbifold.

\section*{Acknowledgments}
The work of K.B. is supported by the EU fourth framework program 
TMR contract FMRX-CT98-0194 (DG 12-MIHT).
This work is also 
supported in part by EU under contracts HPRN-CT-2000-00152 and 
HPRN-CT-2000-00148, in part by IN2P3-CICYT contract Pth 96-3 and 
in part by CICYT, Spain, under contract AEN98-0816.

\appendix 

\section{Embedding of the standard model in  $U(3)\times U(3)$}

We will assume that the model can be embedded in a configuration of D-branes 
of type I strings. In such a case matter fields arise 
as massless fluctuations of open strings stretched between two sets of branes. 
Given $n$ sets of coincident $N_i$, $i=1,\cdots n$,
D-branes, the associated gauge group is 
$U(N_1)\times \cdots \times U(N_n)\equiv SU(N_1)\times \cdots \times 
SU(N_n)\times U(1)_{N_1}\times\cdots \times
U(1)_{N_n}$, with non-abelian gauge couplings $g_{N_i}$ 
and abelian ones normalized as $g_{N_i}/\sqrt{2 N_i}$. 
An open string starting on one of the 
$N_i$ and ending on one of the $N_j$ branes transforms in the representation
$(N_i,{\bar {N_j}})$ of $SU(N_i)\times SU(N_j)_{(1,-1)}$ where $(1,-1)$ are
the $U(1)_{N_i} \times U(1)_{N_j}$ charges. 

For the purpose of embedding the standard model, we choose 
$n=2$ and $N_1 = N_2 =3$, so that
the gauge group is $U(3)_3\times U(3)_2 \equiv 
SU(3)_c\times SU(3)_w \times U(1)_3\times U(1)_2$. 
We denote by $\Q_3$ and 
$\Q_2$ the charges associated to $U(1)_3$ and $U(1)_2$, respectively.
The weak $SU(3)_w$ contains $SU(2)_w \times U(1)_1$
as its maximal subgroup, with the generator $\Q_1$ of $U(1)_1$  
represented in the adjoint of $SU(3)_w$ as $\sqrt{3} \lambda_8$.
Here $\lambda_8$ is the diagonal 
Gell-Mann matrix with entries  $\{1/\sqrt{3},1\sqrt{3},-2\sqrt{3}\}$.

The standard model hypercharge is a linear combination 
of the three $U(1)$ charges $\Q_1$,  $\Q_2$ and $\Q_3$:
\bea
\Q_Y &=& c_1 \Q_1 +c_2 \Q_2 + c_3 \Q_3  
\label{hypeone}
\eea
where the coefficients $c_i$ are such that it reproduces the
standard model representation quantum numbers.

First, note that the Higgs doublets arising from the 
decomposition of the adjoint of $SU(3)_w$ in irreducible representations 
of $SU(2)_w \times U(1)_1$
are not charged with respect to either $\Q_2$ or $\Q_3$.  With 
their hypercharge normalized as $\pm 1/2$ we obtain $c_1=1/6$. Next,
we consider the 
lepton doublets $l$ to arise from the representation 
$({\bf 1},{\bf 3})_L$, while the singlet $e_R$ belongs to the mirror
representation $({\bf 1},{\bf 3})_R $. In order to obtain the 
correct normalization of the corresponding hypercharges, we are led to
$c_2=-2/3$. Finally, for the quark representations
we find two possible choices, 
corresponding to put either the $u^c$ or the $d^c$ quark with 
the quark doublet in the bifundamental representation 
of $SU(3)_c\times SU(3)_w$. 
The first choice leads to the model described in section 4. 
The other choice leads to $c_3=2/3$ with matter representations:
\bea
L_{1,2,3} =({\bf 1},{\bf 3})^+_{(0, 1)}, \qquad 
U_{1,2,3} =({\bf { 3 }}, {\bf 1})^+_{(1,0)}, 
\\
Q_{1} =({\bf 3},{\bf  3 })^-_{(1,1)} \qquad 
Q_{2} =({\bf{  3}},{\bf {3 }})^-_{(1,1)} \qquad 
Q_3 =({\bf { 3 }}, {\bf { 3}})^+_{(1,1)} 
\label{repr2}
\eea
The standard model 
representations are obtained through a $\mathbb{Z}_2$ orbifold on 
the representations $r_f$ as: 
\bea
\begin{array}{lllll} r_f \rightarrow \Theta  r_f:&  
 ({\bf 1},{\bf 3})_L & ({\bf 3},{\bf 3})_L\nonumber \\
r_f \rightarrow -\Theta r_f: & ({\bf 1},{\bf 3})_R & ({\bf 3},{\bf 3})_R
\nonumber \\
r_f \rightarrow  r_f:
&({\bf 3},{\bf 1})_R &&& \nonumber \\
r_f \rightarrow - r_f:
 & ({\bf 3},{\bf 1})_L & &&
\end{array}
\eea
which keeps the standard model fermions and projects the mirrors away.
Only one linear combination is anomaly free and corresponds to the 
hypercharge
\bea
\Q_Y &=&\frac{\Q_1}{{6}} -  \frac{2 \Q_2}{3} +\frac{2 \Q_3}{3}  
\label{hypercharge1}
\eea
The corresponding tree-level gauge coupling is given by:
\bea
\frac{1}{g_Y^2}= \frac{3}{g^2}+ \frac{8}{3}\frac{1}{g_3^2}
\eea
and corresponds to a weak mixing angle $\theta_w$ given by:
\bea
\sin^2{\theta_w}=\frac{1}{4 + \frac{8}{3}\frac{g^2}{g_3^2}}
\eea

Note that both this model and the one presented in section 4 
require the presence of a new brane where the open strings giving rise to 
$L$ and $D^c$ or $U$ will end. One way to avoid the
introduction of the new brane is to make use of the fact that
the representation ${\bf \bar{3}}$ can be obtained as the 
antisymmetric product of two ${\bf 3}$'s.   
$L$ and $D^c$ can then be identified with massless
exitations of open strings with both ends on the weak and color 
D-branes, respectively, and corresponding $U(1)$ charges,
$L=({\bf 1},{\bf 3})_{(0, 2)}$ and
$D^c=({\bf {\bar 3 }}, {\bf 1})_{(-2,0)}$. The hypercharge generator is then:
\bea
\Q_Y &=&\frac{\Q_1}{{6}} +  \frac{ \Q_2}{3} -\frac{ \Q_3}{3}  
\label{hypercharge2}
\eea
%


\begin{thebibliography}{99}
%
\bibitem{ABQ1}
I.~Antoniadis, K.~Benakli and M.~Quir\'os,
\NPB{583}{2000}{35}.
%
\bibitem{Hatanaka}
Y.~Hosotani,
 \PLB{126}{1983}{309};
H.~Hatanaka, T.~Inami and C.~S.~Lim,
\MPL{13}{1998}{2601};
H.~Hatanaka,
\PTP{102}{1999}{407};
G.~Dvali, S.~Randjbar-Daemi and R.~Tabbash,
\texttt{hep-ph/0102307}.
N.~Arkani-Hamed, A.~G.~Cohen and H.~Georgi,
\texttt{hep-ph/0105239}.
A.~Masiero, C.A.~Scrucca, M.~Serone and L.~Silvestrini,
\texttt{hep-ph/0107201}.


\bibitem{A} 
I.~Antoniadis,
\PLB{246}{1990}{377}.
 
\bibitem{AB} I.~Antoniadis and K.~Benakli, \PLB{326}{1994}{69};
K.~Benakli, \PLB{386}{1996}{106}; 
C.~Bachas, \texttt{hep-th/9509067}.


\bibitem{savas} 
I. Antoniadis, S. Dimopoulos, A. Pomarol and M. Quir\'os,  
\NPB{554}{1999}{503}.

\bibitem{Delgado}
A.~Delgado, A.~Pomarol and M.~Quir\'os,
\PRD{60}{1999}{095008}.

\bibitem{BHN} R. Barbieri, L. Hall and Y. Nomura, 
\PRD{63}{2001}{105007};
N.~Arkani-Hamed, L.~Hall, Y.~Nomura, D.~Smith and N.~Weiner,
\NPB{605}{2001}{81};
A.~Delgado and M.~Quir\'os,
\NPB{607}{2001}{99}.

\bibitem{gero}
A.~Delgado, G.~von Gersdorff, P. John and M.~Quir\'os, 
\texttt{hep-ph/0104112};
%
R.~Contino and L.~Pilo, \texttt{hep-ph/0104130};
%
Y.~Nomura, \texttt{hep-ph/0105113};
%
N.~Weiner, \texttt{hep-ph/0106021};
%
A.~Delgado, G.~von Gersdorff and M.~Quir\'os, \texttt{hep-ph/0107233}.

\bibitem{ADD}
I.~Antoniadis, S.~Dimopoulos and G.~Dvali,
\NPB{516}{1998}{70}.

\bibitem{Fayet}
P.~Fayet,
\PLB{159}{1985}{121}.

\bibitem{AKT}
I.~Antoniadis, E.~Kiritsis and T.~N.~Tomaras,
\PLB{486}{2000}{186}.

\bibitem{Hunter} F.J.~Gunion, H.E.~Haber, G.~Kane and S.~Dawson, {\it The 
Higgs Hunter Guide} (Addison-Wesley Pub. Company, Redwood City, CA, 1990).

\bibitem{GS} M.~B.~Green~  and J.~H.~Schwarz,
\PLB { 149} {1984} {117};
A.~Sagnotti, \PLB {294} {1992} {196}.

\bibitem{DSW} M.~Dine, N.~Seiberg and E.~Witten,
\NPB{ 289} {1987} {589};
L.~E.~Ibanez, R.~Rabadan and A.~M.~Uranga,
\NPB {542} {1999} {112}.

\end{thebibliography}
\end{document}